\documentclass[twocolumns,traditabstract]{aa}  
\usepackage{changepage}
\usepackage{tabularx,multirow,booktabs,blindtext}
\usepackage{graphicx}
\usepackage{subfig}
\usepackage{placeins} 
\usepackage{txfonts}
\usepackage{graphicx,rotating}
\usepackage{subfig}
\usepackage{adjustbox}
\usepackage{txfonts}
\usepackage[T1]{fontenc}
\usepackage{epsfig}
\usepackage{natbib}
\usepackage{color}
\usepackage{amssymb}  
\usepackage{tabularx}
\usepackage{comment}
\usepackage{longtable}
\usepackage{ltablex}
\usepackage{caption, booktabs}

\DeclareGraphicsExtensions{.pdf,.png,.jpg}
\bibpunct{(}{)}{;}{a}{}{,}

\begin{document} 

\authorrunning{Castignani et al.}
\titlerunning{Molecular gas in cool-core brightest cluster galaxies at $z\simeq0.4$}

\title{Molecular gas in cool-core brightest cluster galaxies at $z\simeq0.4$}

   \author{
   G. Castignani\inst{1}
   \and
   F. Combes\inst{2,3}
   \and
   P. Salomé\inst{2}
   \and
   A. Edge\inst{4}
   \and
   P. Jablonka\inst{5,2}
   }

   \institute{INAF - Osservatorio  di  Astrofisica  e  Scienza  dello  Spazio di  Bologna,  via  Gobetti  93/3,  I-40129,  Bologna,  Italy \\ e-mail: \texttt{gianluca.castignani@inaf.it} 
            \and
Observatoire de Paris, PSL University, Sorbonne University, CNRS, LUX, 75014 Paris, France
\and
Coll\`ege de France, 11 place Marcelin Berthelot, 75231 Paris, France
\and
Centre for Extragalactic Astronomy, Department of Physics, Durham University, Durham DH1 3LE, UK
\and
Laboratoire d’astrophysique, École Polytechnique Fédérale de Lausanne (EPFL), CH-1290 Versoix, Switzerland}

                \date{Received 4 December 2024 / Accepted 14 April 2025}

  \abstract
  {The brightest cluster galaxies (BCGs) of today are passive and very massive galaxies at the center of their clusters. They still accrete mass by swallowing companions and flows of cold gas, processes which are regulated by radio-mode active galactic nucleus (AGN) feedback. However, the formation history of BCGs is still a matter of debate. We report new findings based on millimeter observations performed with the Northern Extended Millimeter Array (NOEMA) interferometer mapping the cold molecular gas (CO) that feeds the star formation of distant BCGs. We selected three sources among the strongest cool-core BCGs at intermediate redshifts ($z\simeq0.4$), namely, RX 1532, MACS 1447, and CHIPS 1911. Previous unresolved millimeter observations and multi-wavelength analyses have shown that they are among the most star forming (${\rm SFR}\simeq100~ M_\odot/{\rm yr}$) and gas rich ($M_{H_2}\simeq10^{11}~M_\odot$) BCGs at intermediate redshifts. The selected sources are thus caught in a phase of rapid mass assembly, which makes them ideal targets for high-resolution observations of their molecular gas. We find that all three BCGs show point-like and steep-spectrum continuum emission at millimeter wavelengths, with a spectral slope of $\alpha=0.6$, which we interpret as being optically thin synchrotron emission from the AGN.  By combining our NOEMA intensity and velocity maps with archival deep optical/infrared images from the {\it Hubble Space Telescope}, we unambiguously detected in situ star formation, filaments of accreting cold gas likely regulated by AGN feedback, a disturbed morphology associated with tidal tails of molecular gas, and gas compression and tails originating from the {stripping of gas.} {While effective condensation of the intracluster medium is required to explain the large molecular gas reservoirs, the BCGs exhibit a broad variety of environment-driven mechanisms responsible for the processing of their cold gas: flows of cooling gas (RX~1532), ram pressure or sloshing of the intracluster medium} (MACS~1447), and galactic tides (CHIPS~1911). This study thus provides new insights into the physical mechanisms responsible for the mass assembly of galaxies hosting AGN at the center of clusters. }


   \maketitle

\begin{table*}[]
\caption{Properties of our targets.}
\centering
\begin{adjustwidth}{-0.5cm}{}
\begin{center}
\begin{tabular}{ccccccccc}
\hline\hline\\
Galaxy  & R.A. & Dec. & $z_{spec}$ &  $M_\star$ & SFR & sSFR & sSFR$_{\rm MS}$ & $K_0$  \\
   & (hh:mm:ss.s) & (dd:mm:ss.s) &  &  ($10^{11}~M_\odot$) & ($M_\odot$/yr) & (Gyr$^{-1}$) & (Gyr$^{-1}$) & (keV~cm$^2$) \\ 
  (1) & (2) & (3) & (4) & (5) & (6) & (7) & (8) & (9) \\\\
 \hline
 \\
 RX~1532 & 15:32:53.79 & +30:20:59.5    & 0.361 &  $1.34\pm0.03$  & $59\pm2$~(UV) & 0.44$\pm$0.02  & 0.08  & $17\pm2$\\
        &             &               &         &                 & $85\pm24$~(H$\alpha$) & 0.64$\pm$0.18 &  & \\\\

\hline  \\
MACS~1447 & 14:47:26.02 & +08:28:25.1 & 0.376 & 6.3 & 61.3~(H$\alpha$)  &  0.10 & 0.05 & $12\pm2$ \\
 &  &  &  &  & 78.5~([O~II]) & 0.12 \\\\
\hline\\
CHIPS~1911 & 19:11:02.00 &	+44:55:20.0 & 	0.485 & $2.21^{+1.02}_{-0.65}$ & $143^{+31}_{-26}$~(mid-IR) & $0.64^{+0.23}_{-0.32}$ & 0.10 & $17^{+2}_{-9}$ \\\\ 
 &  &	 & 	 &  & $115^{+15}_{-13}$~([O~II]) & $0.52^{+0.17}_{-0.25}$  &  &  \\\\ 
\hline 
\end{tabular}
 \end{center}
 \tablefoot{Column description: (1) BCG name; (2-3) J2000 equatorial coordinates; (4) spectroscopic redshift; (5) stellar mass from 
\citet[][for RX~1532]{Burke2015}, 
from \citet[][for MACS~1477]{Liu2012}, and from this work (for CHIPS~1911); (6) UV, H$\alpha$, [O~II], and mid-IR based SFRs of RX~1532 \citep{Fogarty2015},  MACS~1447 \citep{Liu2012}, and CHIPS~1911 \citep{Somboonpanyakul2021a}, corrected for dust extinction; (7) specific SFR determined as ${\rm sSFR}={\rm SFR}/M_\star$; (8) sSFR for main sequence field galaxies with redshift and stellar mass of our targets estimated using the relation found by \citet{Speagle2014}; (9) central entropy $K_0=k_B~T_X~n_e^{-2/3}$ of the cluster estimated using X-ray {\it Chandra} observations by \citet{Donahue2015} for RX~1532, \citet{Somboonpanyakul2021a} for CHIPS~1911, and \citet{PrasowEmond2020} for MACS~1447.}
\label{tab:BCG_properties}
\end{adjustwidth}
\end{table*}

\section{Introduction}
Galaxy clusters are very dense structures where most of the baryonic mass is in the form of very hot gas with temperatures of $\sim10^7$~K. X-ray observations have revealed that the center of some of these clusters produces intense radiation, showing that the intracluster medium (ICM) is cooling radiatively \citep[e.g.,][for a review]{Fabian1994}. A simple calculation of the cooling time, on the order of $\sim300$~Myr, which is much smaller than the Hubble time, leads to a high accretion rate of $\sim(100-1000)~M_\odot$/yr of cold gas, which has never been observed in the cluster cores, either as cold gas or young stars.

Brightest cluster galaxies (BCGs) live at the center of galaxy clusters, and they are optimal laboratories to study the effect of the cluster environment on galaxy evolution.  There exists a complex co-evolution between the active galactic nucleus (AGN) activity of BCGs, the growth of super massive black holes at the center of galaxies, the galaxy stellar content, and star formation history, as well as the megaparsec-scale environments
\citep{Heckman2014,Burke2015,Fogarty2015,Somboonpanyakul2022,Poitevineau2023}.

In the local Universe, BCGs are commonly associated with passively evolving massive cD ellipticals at the center of cluster cores \citep{Lauer2014}, and they often host radio galaxies \citep{Zirbel1996}. The BCGs with powerful radio jets are in fact almost always observed in the center of cool-core clusters \citep{Sun2009}. BCG jets are able to release energy likely via sound waves into the ICM out to the cluster periphery, thus preventing the formation of the above-mentioned strong cooling flows \citep[e.g.,][for a review]{Fabian2012}.  
Deep X-ray images have even shown that the radio jets create bubbles in the hot ICM, turbulence, and shocks. Altogether, these phenomena slow down the cooling of gas  \citep[e.g.,][]{McNamara_Nulsen2007,Walker2017}. As occurs in the prototypical Perseus cluster, the gas therefore cools far from the center at the borders of the cavities sculpted in the ICM by the AGN jets
\citep{Salome2006}.

Altogether, such a self-regulating mechanism, known as radio-mode AGN feedback \citep[][]{Fabian2012}, not only solves the cooling-flow problem, but it also explains why large reservoirs of molecular gas and high star formation activity are rarely observed
in the BCGs, which require effective ICM condensation. To explain their high masses and large radii observed at $z\sim0$, dry mergers following the galaxy formation at high-redshift $z\sim5$ \citep{DeLucia_Blaizot2007} are invoked as the common evolutionary path for BCGs. However, a large variety of environmental processes are still viable to explain the mass assembly and star formation history of BCGs: dynamical friction \citep{White1976}, galactic cannibalism \citep{Hausman_Ostriker1978}, and interactions with the ICM \citep{Stott2012}. 

{A number of studies have investigated the molecular gas properties of distant clusters, suggesting the ubiquitous presence of large molecular gas reservoirs in and around $z\gtrsim1$ BCGs \citep{Webb2017,Castignani2018,Castignani2020b,Dunne2021,Barfety2022,Castignani2022a}. At lower redshifts, studies of local and intermediate-$z$ BCGs in cool-core clusters have found molecular gas that fuels star formation \citep[e.g.,][]{Edge2001,Salome_Combes2003,Odea2008,Hamer2012,McDonald2012,Rawle2012,McDonald2014,Russell2014,Tremblay2016,Russell2019}, with total masses that are typically around one to two orders of magnitude lower than those theoretically predicted from the X-ray cooling rates.} These flows of cold gas in the BCGs are common in dynamically relaxed cool-core clusters \citep{Fabian2006,Eckert2011}, which suggests that large-scale flows of gas regulate the star formation and are often moderated or suppressed by AGN feedback.  A recent compilation of a large sample of cool-core clusters and their gas maps \citep{Olivares2019} has also showed that the filamentary multi-phase structures (H$\alpha$, CO) are ubiquitous in nearby BCGs and that the gas is in-falling with very low velocities, which is compatible with the chaotic cold accretion model \citep{Gaspari2015}.

These filamentary structures around BCGs result from the local cooling of the ICM via thermal instabilities \citep{Beckmann2019} that are likely triggered by the AGN jets. Filaments of cold gas exhibit a large range in morphology and cooling-to-free-fall time ratio, which reflects the ICM inhomogeneities and the complexity of the AGN feedback. Nevertheless, the morphology, distribution (disk versus filament proportion in molecular gas), and the underlying gas processing mechanisms at play regulated by AGN feedback are still unknown for distant BCGs.

{The present work is aimed at answering the open questions related to the mass assembly as well as the stellar and star formation properties of BCGs by means of} high-resolution maps of cold molecular gas in intermediate-$z$ BCGs taken with the Northern Extended Millimeter Array (NOEMA) interferometer. We focus on three star forming -- star formation rates in the range $\simeq(60-140)~M_\odot$/yr -- intermediate-redshift ($z\sim0.4-0.5$) BCGs of cool-core clusters. Thus, our goal is to probe the physical processes responsible for the gas feeding the star formation in intermediate-$z$ counterparts of local star-forming BCGs
{($\gtrsim10~M_\odot$~yr$^{-1}$) such as the famous Perseus~A and Cygnus~A \citep[e.g.,][]{Salome2011,Privon2012,Fraser-McKelvie2014}.} 

This work is part of a series of papers whose goal is to investigate the star formation and mass assembly of BCGs over cosmic time, mainly by exploiting targeted observations at millimeter wavelengths of their molecular gas, using both single-dish \citep{Castignani2019,Castignani2020c,Castignani2020a,Castignani2022b,Castignani2023} and interferometry \citep{Castignani2020b,Castignani2022a}. The paper is organized as follows. In Sect.~\ref{sec:BCGsample}, we describe the properties of our target BCGs. In Sect.~\ref{sec:mm_observations}, we describe the millimeter observations and data analysis. In Sect.~\ref{sec:results}, we report our results;  in Sect.~\ref{sec:discussion}, we discuss them; and in Sect.~\ref{sec:conclusions}, we draw our conclusions. Throughout this work, we adopt a flat $\Lambda \rm CDM$ cosmology with a matter density of $\Omega_{\rm m} = 0.30$, a dark energy density of $\Omega_{\Lambda} = 0.70$, and the Hubble constant $h=H_0/100\, \rm km\,s^{-1}\,Mpc^{-1} = 0.70$.
The stellar masses and star formation rates adopted in this work rely on the \citet{Chabrier2003} initial mass function (IMF).

\begin{figure*}[h!]
\centering
\includegraphics[trim={0cm 0cm 5cm 0cm},
width=0.45\textwidth]{{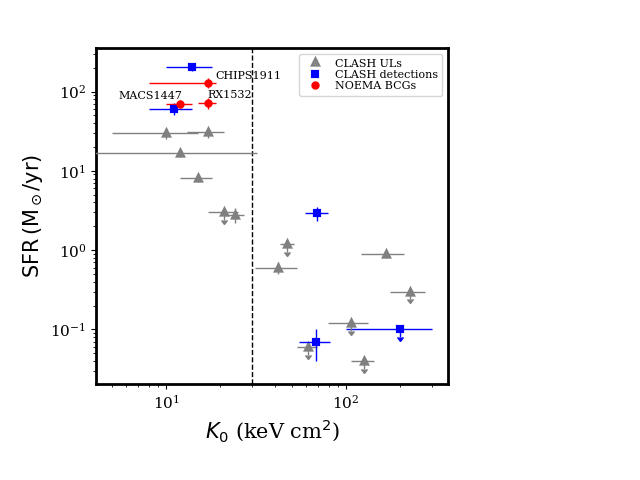}}\hspace{0.5cm}
\includegraphics[trim={0cm 0cm 5cm 0cm},width=0.45\textwidth]{{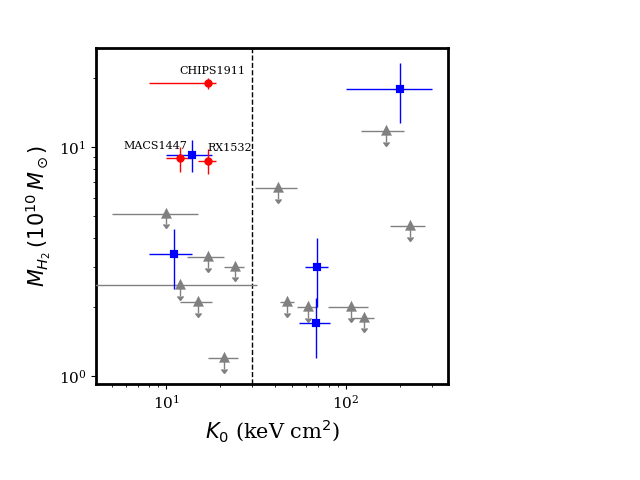}}\hspace{0.5cm}
\caption{Star formation rate (left) and molecular gas mass (right) versus core entropy for intermediate-redshift BCGs with existing CO observations from the CLASH survey and this work. The BCGs RX~1532, MACS~1447, and CHIPS~1911, which are the subject of this work, are highlighted as red circles. With blue squares, we highlight CLASH BCGs with clear CO detections from the IRAM~30m telescope \citep[MACS~0329, Abell~1423, MACS~1206, MACS~2129,][]{Castignani2020a} and ALMA \citep[MACS~1932,][]{Fogarty2019}. CLASH BCGs with upper limits to their total $H_2$ gas mass by \citet{Castignani2020a} are reported as gray triangles. We refer to the text for further details. The vertical dashed line is located at $K_0 = 30$~keV~cm$^{2}$.}
\label{fig:MH2_SFR_vs_entropy}
\end{figure*}

\section{The sample of cool-core BCGs at $z\simeq0.4$} \label{sec:BCGsample}
We wanted to resolve with NOEMA the CO emission from distant star forming BCGs in order to reveal possible asymmetries, the total extent, and the velocity field of the cold gas both in the rotating star forming disk and in the cooling filaments infalling onto the BCGs. 
Therefore, we selected three among the most star-forming and cool-core BCGs at intermediate redshifts. {These are the spectroscopically confirmed RXC~J1532.8+3020   ($z=0.361$), MACS~J1447.4+0828  ($z=0.376$), and CHIPS~J1911.0+4455 ($z=0.485$) BCGs, hereafter denoted as RX~1532, MACS~1447, and CHIPS~1911, respectively, ordered by increasing redshift.} 
At the redshift of RX~1532, MACS~1447, and CHIPS~1911 BCGs, $1^{\prime\prime}$ corresponds to 5.04, 5.17, and  6.00~kpc (physical units), respectively. The selected BCGs have a wealth of ancillary data, both in spectroscopy and in photometry, which enabled the characterization of their stellar and star formation properties, as further outlined in the following.

\subsection{RX~1532 ($z=0.361$)}\label{sec:RX1532}
{RX~1532 belongs to an X-ray cluster, which was first presented in the 
{\it ROSAT} Brightest Cluster Sample \citep[BCS,][]{Ebeling1998}. {The total mass of the cluster, within a radius encompassing the matter density 500 times the critical, was estimated to be $M_{\rm 500}=(3.93\pm1.01)\times10^{14}~M_\odot$ \citep[see][]{PlanckCollaboration2014,PennaLima2017}.} The cluster was later included in the} Cluster Lensing And Supernova survey with Hubble (CLASH) survey. High-resolution, multi-band {\it Hubble Space Telescope (HST)} observations \citep{Postman2012a} and deep X-ray monitoring with Chandra \citep{Donahue2016} were performed for CLASH clusters, including RX~1532. Several spectroscopic campaigns also led to the conﬁrmation of a large number of CLASH cluster members \citep[][and references therein]{Caminha2019}, which make CLASH an excellent survey to study the evolution of cluster galaxies, including BCGs \citep{Postman2012b,Annunziatella2014,Annunziatella2016,Burke2015,Yu2018,Durret2019,Fogarty2019,Castignani2020a}, and their interplay with the intra-cluster medium \citep{Donahue2016,Fogarty2017,DeMaio2020,Levitskiy2024}

Interestingly, ultraviolet (UV), H$\alpha$, H$\beta$, [O~III], and [O~II] observations from the {\it HST} confirmed strong ongoing star-formation over 50~kpc in diameter \citep{Fogarty2015}. This can be appreciated from the {\it HST} image and maps of the BCG \citep[see Figs~1 and 3 in][as well as Figs.~\ref{fig:allBCGs_optical_CO} and \ref{fig:RX1532_SF_CO_maps} below]{Fogarty2015}. They display clear evidence of star forming clumps and elongated filaments up to $\sim5^{\prime\prime}$. Previous {\it Chandra} observations of RX~1532 also excluded the presence of a strong X-ray emitting active galactic nucleus (AGN) and suggested a low-entropy
\citep[$K_0 = 17\pm2$~keV~cm$^2$,][]{Donahue2015} cluster core which favors the presence of cold gas reservoirs.  Overall, RX~1532 is among the cool-core BCGs in the CLASH sample with the highest star formation rates ${\rm SFR}\simeq100~ M_\odot/{\rm yr}$, as inferred from both H$\alpha$ and UV emission \citep{Fogarty2015}, together with MACS~1932 ($z = 0.35$) in the southern hemisphere and MACS~0329 ($z=0.450$). Based on the single-dish IRAM~30m \citep{Castignani2020a} and interferometric ALMA observations \citep{Fogarty2019}, RX~1532 and MACS~1932 show clear detections of cold molecular gas in several CO lines, yielding a large cold molecular gas reservoir of $M_{H_2}\simeq9\times10^{10}~M_\odot$ for both BCGs. {However, for RX~1532, we stress that detections in CO(1$\rightarrow$0), CO(3$\rightarrow$2), and CO(4$\rightarrow$3) were first reported  by \citet{Edge2001}.}

\subsection{MACS~1447 ($z=0.376$)}\label{sec:MACS1447}
Similarly to RX~1532, we have selected the distant BCG of MACS~1447 at $z = 0.376$ {\citep{Ebeling2001,Hicks2013}}.
{The total mass of the cluster was estimated to be $M_{\rm 500}= (7.46^{+0.80}_{-0.86})\times10^{14}~M_\odot$ \citep[see][]{Planck2015,PrasowEmond2020}.}
The BCG has unambiguously
high star formation activity, detected both in [O~II] and in H$\alpha$, with a resulting SFR ranging between 60-80~$M_\odot$/yr \citep{Liu2012}. Past IRAM~30m observations (ID:~274-10), which we publish for the first time in this work (see Sect.~\ref{sec:IRAM30m_observations}), reveal a large and strongly excited reservoir of  molecular gas that sustains the high star formation, similarly to RX~1532, with a total mass of $M_{H_2}\simeq9\times10^{10}~M_\odot$ and an excitation ratio of $r_{32}=0.81\pm0.14$.

This scenario is supported by recent observations by \citet{PrasowEmond2020} in X-rays with {\it Chandra}, that clearly demonstrate that MACS~J1447 is a cool-core cluster, with a low entropy ﬂoor of $(12\pm2)$~keV~cm$^2$ and a well deﬁned extended X-ray (0.5-8 keV) emission coming from the ICM. Such a low level of entropy makes MACS~J1447 one of the strongest cool-core clusters known, even when compared to similar well-studied cool-core clusters at intermediate-$z$, such as MACS~1932,
RX~1532, and MACS~0329 \citep[see Sect.~\ref{sec:RX1532} and e.g.,][]{Fogarty2015}. X-ray cavities to the north and south of the BCG are also found by {\it Chandra}, co-spatial with an
elongated radio emission detected with the Jansky Very Large Array \citep[JVLA,][]{PrasowEmond2020}. A plume in X-rays, to the west of the BCG is co-spatial with a ﬁlament detected by {\it HST}. These ﬁndings suggest that the BCG launches jets that
inﬂate into the ICM and moderate the cooling and condensation of gas onto the BCG via radio-mode AGN feedback. Altogether, these features make MACS~J1447 an excellent intermediate-$z$ analog of Perseus, the benchmark cluster for radio-mode AGN feedback in the local Universe \citep{Fabian2006,Salome2006}.

\subsection{CHIPS~1911 ($z=0.485$)}\label{sec:CHIPS1911}

The third BCG considered in this work is CHIPS~1911 ($z=0.485$). It belongs to a recently discovered cool-core X-ray cluster 
\citep{Somboonpanyakul2021a,Somboonpanyakul2021b}. {Based on {\it Chandra } observations in the X-rays, \citet{Somboonpanyakul2021b} estimated a cluster mass $M_{500}=(6.0\pm0.1)\times10^{14}~M_\odot$.} {Interestingly, the cluster is at low galactic latitude ($\textrm{b}\simeq15.6$~deg) and was therefore overlooked by the MAssive Cluster Survey \citep[MACS,][]{Ebeling2001}, which was restricted to $\textrm{|b|}\geq20$~deg.}
{\it Chandra} observations by \citet{Somboonpanyakul2021a} detected a cool-core associated with the cluster, characterized by a low central entropy $K_0=17^{+2}_{-9}$~keV~cm$^2$. The cluster thus revealed itself as one of the strongest cool-core clusters known, as shown also by a short cluster core cooling time $t_{\rm cool}<1$~Gyr of the ICM and by the radial distribution of temperature, featuring a central drop down to $\lesssim4$~keV within 20~kpc from the cluster center. All above cool-core cluster features are typically found in relaxed clusters \citep[e.g., Phoenix,][]{McDonald2019}.

However, CHIPS~1911 cluster looks very special in this
regard, as its X-ray morphology is very asymmetric at both small and large scales from 20 to 200~kpc, suggesting a recent or ongoing merger. Consistently with this picture, the peakiness and asymmetry of CHIPS~1911 was found to be among the most extreme \citep{Somboonpanyakul2021a}. Altogether, it is thought that an interaction between two groups or clusters, leading to a merger, is able to displace the BCG and perturb the cooling of the ICM, providing some heating, although some cold gas has been observed to cool around cavities produced by a sloshing BCG \citep[e.g., Abell~1795,][]{Salome_Combes2004,Russell2017}. Cluster mergers may have the power to destroy cool cores through shocks and mixing, and increase the ICM entropy to values $>~30$~keV~cm$^2$ \citep[e.g.,][]{Poole2008,Donnert2017}. Nonetheless, cool cores have been observed in group-group mergers \citep{OSullivan2019,Pan2020}.

Furthermore, \citet{Somboonpanyakul2021a} found that the X-ray peak of the cluster is displaced to the north of the BCG and it might contain low-entropy gas (see e.g., their Fig.~1 and Fig.~\ref{fig:allBCGs_optical_CO} of this work), suggesting a merger event. CHIPS~1911 BCG has very blue colors, implying high ongoing star formation rate ${\rm SFR}\gtrsim100~M_\odot/{\rm yr}$, as revealed by mid-infrared WISE and [O~II] emission \citep{Somboonpanyakul2021a}. While prior observations of cold molecular gas feeding star formation are missing in the literature, new NOEMA interferometric observations of the CO emission from the BCG detected a large total gas reservoir with $M_{H_2}\simeq2\times10^{11}~M_\odot$ (see Sect.~\ref{sec:results}), comparable to that of the other two BCGs considered in this work.

\subsection{Stellar mass and SFR estimates}\label{sec:Mstar_SFR}
The three BCGs that are the subject of this work have similar stellar masses in the range $\simeq(1-6)\times10^{11}~M_\odot$, consistent with those found for massive cD ellipticals \citep{Zhao2015}. They also have comparable SFR estimates ranging between $\simeq(60-140)~M_\odot$/yr, all corrected for dust attenuation and estimated using UV, H$\alpha$, mid-IR, or [O~II] emission as SFR indicators \citep{Liu2012,Fogarty2015,Somboonpanyakul2021a}. 
The BCGs are thus star forming, with SFR values that are a factor of $\sim2$, for MACS~1447, and $\sim(5-8)$, for RX~1532 and CHIPS~1911, higher than that of the star formation main sequence (MS) at the BCG redshift \citep{Speagle2014}. {For a comparison, only $\sim10\%$ of the BCGs at $z<0.5$ in \citet{Pulido2018} and \citet{Webb2015b} samples have SFR$>50~M_\star$/yr, similarly to the three BCGs of this work.}  Coordinates, stellar masses, star formation properties, and central cluster entropy are summarized in Table~\ref{tab:BCG_properties}.
{We report in the following further details concerning the stellar mass and SFR estimates of the three BCGs.}

\subsubsection{{RX~1532}}
{For RX~1532, stellar mass is from \citet{Burke2015}, who estimated it by performing a fit to the spectral energy distribution (SED) using {\it HST} photometry, evaluated within 50 kpc of the BCG. Dust corrected SFR estimates, based on UV and H$\alpha$ emission,  are instead from \citet{Fogarty2015}. 
As these $M_\star$ and SFR estimates rely 
on the \citet{Salpeter1955} IMF, we have rescaled them to a \citet{Chabrier2003} IMF, multiplying them by factor of 0.61 \citep[e.g.,][]{Madau_Dickinson2014}.}

\subsubsection{{MACS~1447}}
{Stellar mass and SFR estimates of MACS~1447 are taken from \citet{Liu2012}. The latter were estimated using H$\alpha$ and [O~II] emission, after correcting for dust attenuation. Stellar masses were instead estimated by the authors by applying spectral synthesis analysis to the Sloan Digital Sky Survey (SDSS) spectrum of the BCG. Both  $M_\star$ and SFR estimates rely on the \citet{Chabrier2003} IMF, and therefore they have not been rescaled.}

\subsubsection{{CHIPS~1911}}
For CHIPS~1911, \citet{Somboonpanyakul2021a} estimated its SFR based on the [O~II] and mid-IR WISE emission, using the \citet{Kennicutt1998} and \citet{Cluver2017} relations, which rely on the \citet{Salpeter1955} and \citet{Kroupa2002} IMF, respectively. We then rescaled the [O~II]-based SFR estimate to the \citet{Chabrier2003} IMF. Namely, we multiply by a factor of 0.61, while we assumed the conversion between Kroupa and Chabrier IMFs negligible \citep[see e.g.,][]{Madau_Dickinson2014}.

We did not find any stellar mass estimate for CHIPS~1911 BCG in the literature. However, analogously to \citet[][see their Fig.~5]{Somboonpanyakul2021a}, we built the optical-to-infrared spectral energy distribution (SED) of the BCG using \textsf{grizy} optical photometry from Pan-STARRS \citep{Tonry2012,Chambers2016,Flewelling2020}
 and infrared photometry from the W1 and W2 channels of WISE, at 3.4 and 4.6~$\mu$m \citep{Wright2010}.
We then estimated the stellar mass of CHIPS~1911 via modeling of the SED with both {\tt\string LePhare} \citep{Arnouts1999,Ilbert2006} and {\tt\string CIGALE} \citep{Burgarella2005,Noll2009,Boquien2019} using {\tt\string GAZPAR}    online service\footnote{https://gazpar.lam.fr/home}. In both cases, we assumed the \citet{Chabrier2003} IMF.
When using {\tt\string LePhare}, we adopted a library of synthetic spectra generated using the stellar population synthesis model
of \citet{Bruzual_Charlot2003}. Namely, we used the library used in \citet{Ilbert2015} which comprises galaxy templates with exponentially declining star formation history, as in \citet{Ilbert2010}, and two others with delayed star formation. The modeling with {\tt\string LePhare} yields a stellar mass estimate of $M_\star=(2.21^{+1.02}_{-0.65})\times10^{11}~M_\odot$, which we adopt throughout this work, in agreement with the best fit value $M_\star=2.79\times10^{11}~M_\odot$ which is obtained instead with {\tt\string CIGALE}.

\subsection{Molecular gas, star formation, and cool cores}
We now discuss the global star formation and cold gas properties of the three BCGs in connection with the central entropy and in comparison with intermediate-$z$ BCGs from the literature. Figure~\ref{fig:MH2_SFR_vs_entropy} displays the SFR (left) and the total $H_2$ gas mass (right) of the BCGs RX~1532, MACS~1447, and CHIPS~1911, that are plotted as a  function of the entropy level of the cluster core. Cold $H_2$ gas masses are from IRAM~30m observations in the cases of RX~1532 \citep{Castignani2020a} and MACS~1447 (this work, see Sect.~\ref{sec:IRAM30m_observations}) and from NOEMA (this work, see Sect.~\ref{sec:NOEMAobservations}). We refer to the following sections for further details. For a comparison, we also included additional intermediate-$z$ BCGs from the CLASH sample with available CO observations from the literature \citep{Castignani2020a,Fogarty2019}. We also distinguish between those BCGs with CO detections (blue squares) from those with only upper limits in CO (gray triangles) and thus in the $H_2$ gas mass. To allow a homogeneous comparison, for all gas masses that are plotted in the figure, we assumed a Galactic CO-to-$H_2$ conversion factor $\alpha_{\rm CO}=4.36~M_\odot~({\rm K}~{\rm km}~{\rm s}^{-1}~{\rm pc}^2)^{-1}$.

The three BCGs that are the subject of this work clearly populate the regions with the highest SFR and $M_{H_2}$ values, as well as the lowest entropy levels, well below the threshold of $K_0=30$~keV~cm$^2$ (vertical solid dash line), which fiducially defines cool-core clusters.
They are thus among the most star forming, gas-rich, and strongest cool-core BCGs at intermediate-$z$, together with the CLASH BCGs MACS~1932 \citep{Fogarty2019} and  MACS~0329 \citep{Castignani2020a}, also included in Fig.~\ref{fig:MH2_SFR_vs_entropy}, and the cool-core and gas-rich BCG of Phoenix cluster \citep[$z=0.597$,][]{McDonald2014,McDonald2019}.
Phoenix BCG is an extreme outlier among these cool-core BCGs at intermediate-$z$. Therefore, we prefer not to show it in the figure. Indeed, it has an extremely low central entropy $\sim2$~keV~cm$^2$ \citep{McDonald2019}, and it is characterized by an exceptional starburst phase \citep[SFR~$\simeq800~M_\odot$/yr,][]{McDonald2013} sustained by a large molecular gas reservoir with $M_{H_2}\simeq2\times10^{10}~M_\odot$ \citep{McDonald2014}. The present work thus aims to map such large molecular gas reservoirs in three of these cool-core BCGs in order to probe the physical mechanisms responsible for their stellar mass assembly and gas fueling, as described in the following sections.

\section{Millimeter observations and data reduction}\label{sec:mm_observations}

In this section we describe the millimeter observations of the three BCGs and the corresponding data reduction.

\begin{figure}[h!]
\centering
\includegraphics[width=0.4\textwidth]{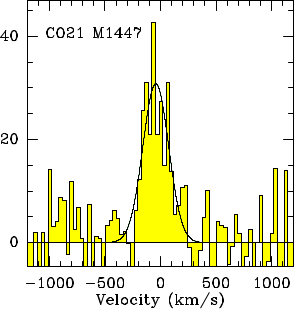}\\\vspace{0.3cm}
\includegraphics[width=0.4\textwidth]{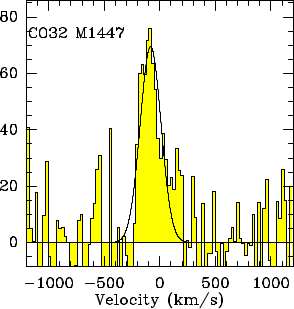}
\caption{Baseline-subtracted  CO(2$\rightarrow$1) and CO(3$\rightarrow$2) spectra of MACS~1447 obtained with IRAM~30m observations. The solid lines show the Gaussian best ﬁts. The ﬂux (y-axis) in units of mJy is plotted against the relative velocity with respect to the BCG redshift, as in Table~\ref{tab:BCG_properties}.}
\label{fig:COspectra_IRAM30m_MACS1447}
\end{figure}

\begin{figure*}[h!]
\includegraphics[width=0.3\textwidth]{{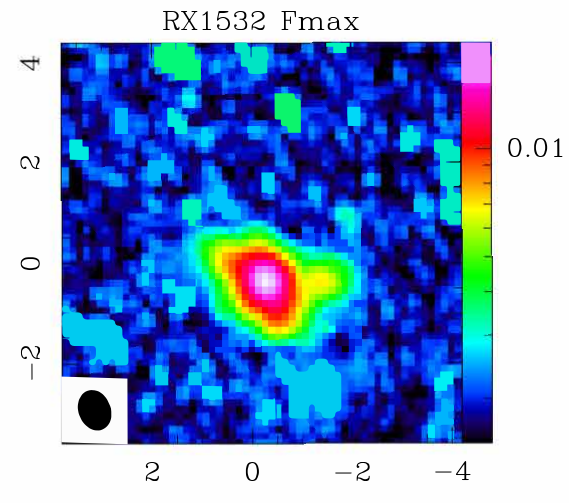}}\hspace{0.5cm}
\includegraphics[width=0.32\textwidth]{{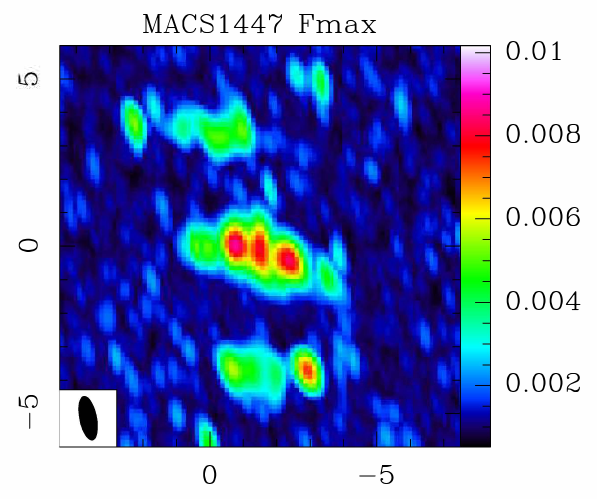}}\hspace{0.5cm}
\includegraphics[width=0.32\textwidth]{{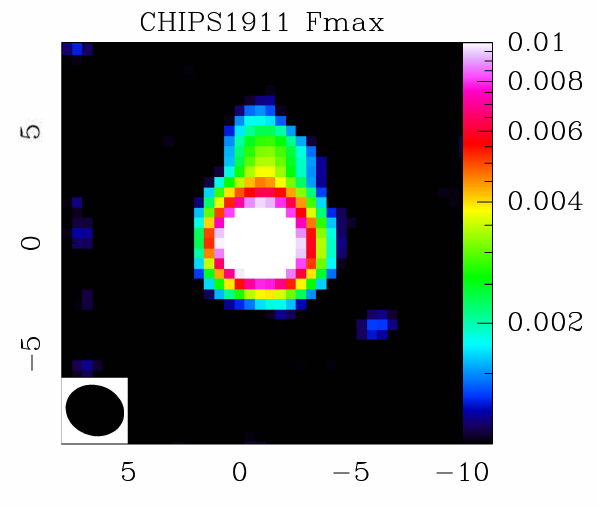}}\vspace{0cm}
\\
\includegraphics[width=0.3\textwidth]{{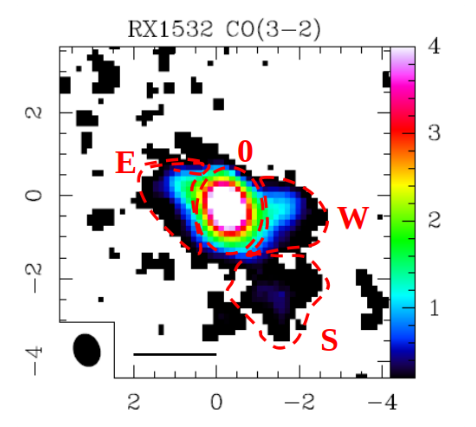}}\hspace{0.5cm}
\includegraphics[width=0.3\textwidth]{{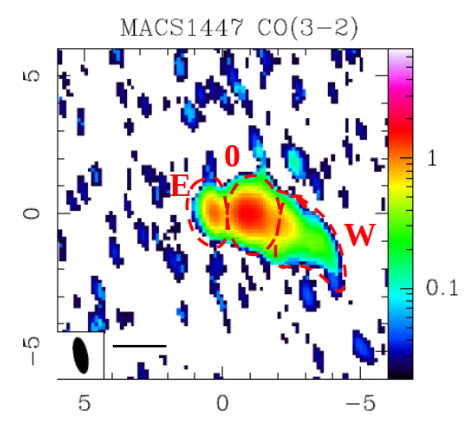}}\hspace{0.5cm}
\includegraphics[width=0.3\textwidth]{{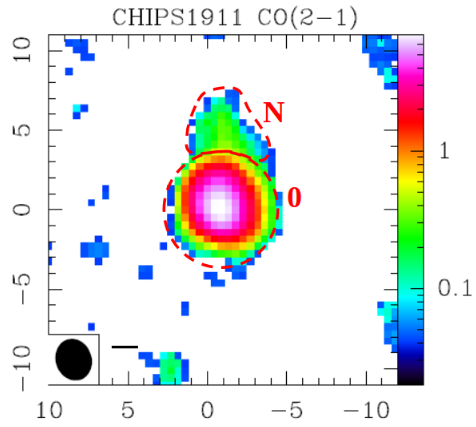}}\vspace{0cm}
\\
\includegraphics[width=0.32\textwidth]{{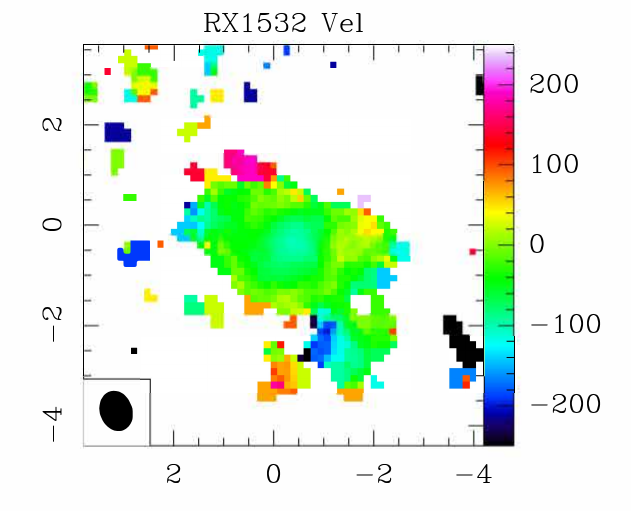}}\hspace{0.5cm}
\includegraphics[width=0.3\textwidth]{{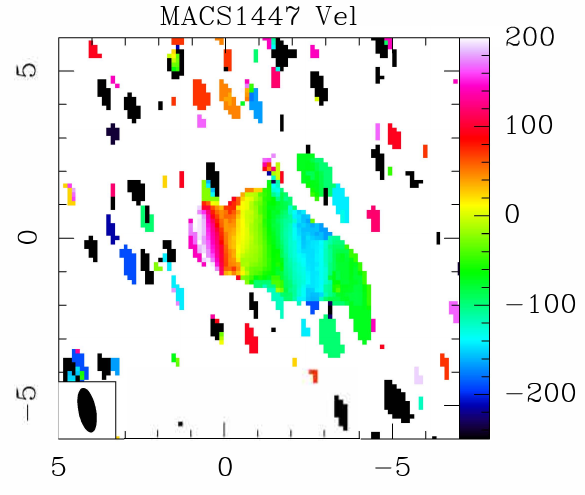}}\hspace{0.5cm}
\includegraphics[width=0.3\textwidth]{{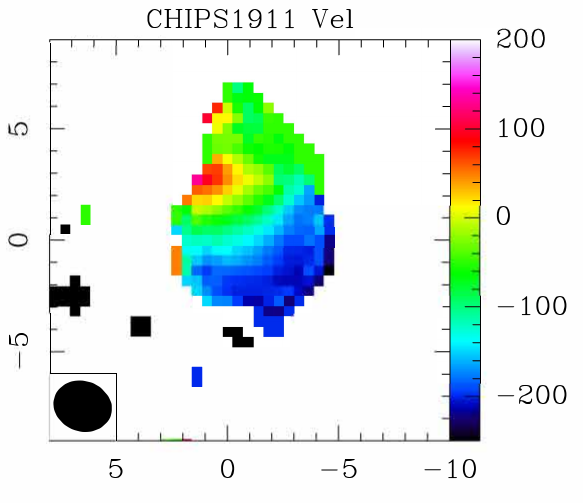}}\vspace{0cm}
\\
\includegraphics[width=0.3\textwidth]{{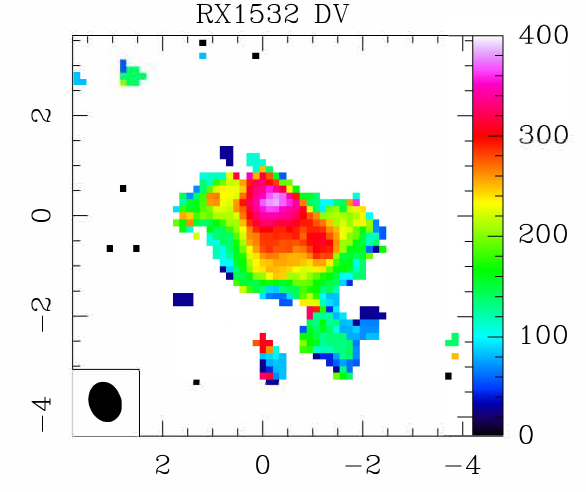}}\hspace{0.5cm}
\includegraphics[width=0.3\textwidth]{{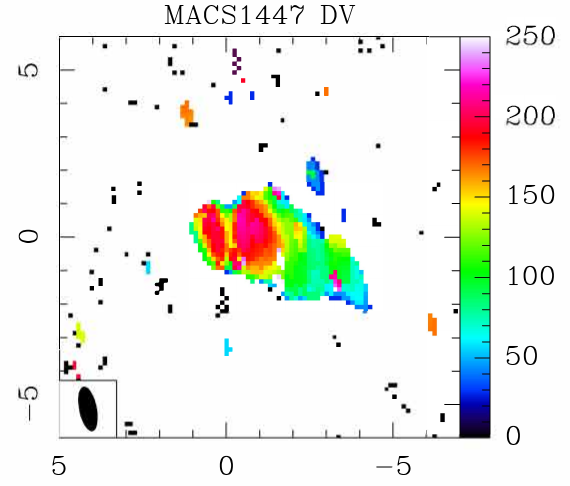}}\hspace{0.5cm}
\includegraphics[width=0.3\textwidth]{{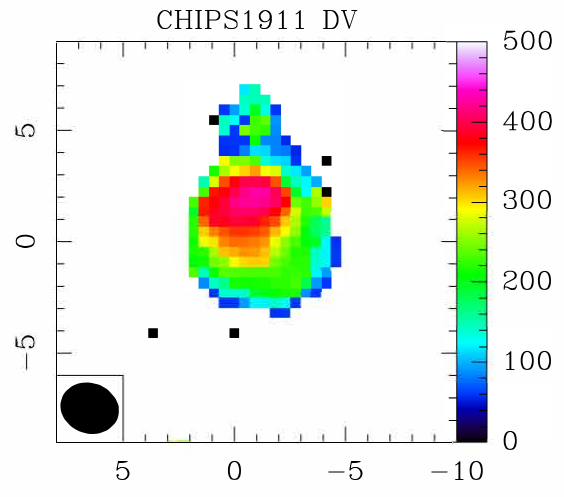}}\vspace{0.1cm}
\caption{From left to right: NOEMA maps of RX~1532 (left), MACS~1447 (center), and CHIPS~1911 (right). From top to bottom: Maps of maxima, evaluated at each pixel of each map; moment-0 maps; moment-1 maps; and moment-2 maps. Coordinates are reported as separation from the phase centers of the observations, as reported in Table~\ref{tab:BCG_properties}. In each panel the NOEMA beam size is reported at bottom left, while
the velocity range used for the integration is between -500 and +500 km~s$^{-1}$ for each map. {In each of the moment-0 maps, the solid black line at the bottom left denotes a scale of 10~kpc at the BCG redshift.} Sub-regions denoted with the labels 0, N, S, W, and E, which were used to extract spectra, are defined by the red dashed lines in the moment-0 maps. We refer to the text for further details. From top to bottom: Values in the color bars correspond to units of Jy$\cdot$beam$^{-1}$ (maps of maxima), Jy$\cdot$km~s$^{-1}$ (moment-0 maps), and km~s$^{-1}$ (moment-1 and moment-2 maps), where velocities are evaluated with respect to the BCG redshifts, as reported in Table~\ref{tab:BCG_properties}.}
\label{fig:NOEMAmaps}
\end{figure*}

\begin{figure*}[h!]
\centering
\includegraphics[width=0.32\textwidth]{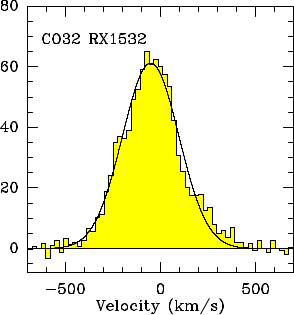}\hspace{0.2cm}
\includegraphics[width=0.32\textwidth]{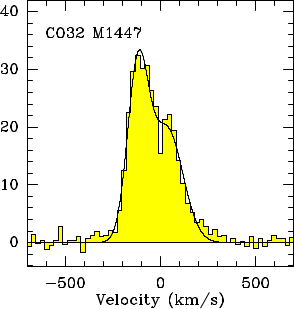}\hspace{0.2cm}
\includegraphics[width=0.32\textwidth]{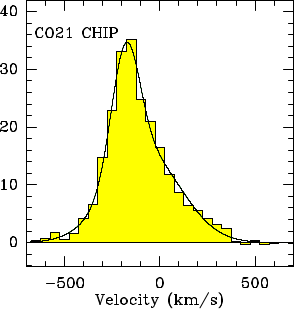}\\\vspace{0.5cm}
\caption{Spectra of RX~1532, MACS~1447, and CHIPS~1911 BCGs obtained with NOEMA. For each BCG, we considered the total emission component (see Table~\ref{tab:properties_mol_gas}).
The solid lines show the continuum-subtracted best ﬁts to the CO emission, where two Gaussian curves are used for MACS~1447 and CHIPS~1911, and one Gaussian only for RX~1532.
The ﬂux (y-axis) in units of mJy, is plotted against the relative velocity with respect to the BCG redshift, as in Table~\ref{tab:BCG_properties}.}
\label{fig:COspectra_total}
\end{figure*}

\subsection{IRAM~30m observations of MACS~1447}\label{sec:IRAM30m_observations}

We observed MACS~1447 BCG using the Eight Mixer Receiver (EMIR) mounted onto the IRAM~30m telescope at Pico Veleta in Spain. The observations were carried out in April 2011 (ID: 274-10; PI: A.~Edge). Each of the E090, E150, and E230 receivers, operating between 1 and 3~mm, offers 4$\times$4~GHz instantaneous bandwidth covered by the correlators.

We used the E090 and E230 receivers, to simultaneously target the CO(1$\rightarrow$0) and CO(3$\rightarrow$2) lines,
redshifted to 3.6~mm and 1.2~mm, respectively.
In addition, we observed the source with the E090 and E150 receivers, simultaneously,
to target the CO(1$\rightarrow$0) and CO(2$\rightarrow$1) lines, the latter being redshifted to 1.8~mm.

The IRAM~30m half-power beam width (HPBW) is $\sim16$~arcsec~$\frac{\lambda_{\rm obs}}{2~{\rm mm}}$ \citep{Kramer2013}, where $\lambda_{\rm obs}$ is the observer frame wavelength. MACS~1447 is thus unresolved by our IRAM~30m observations. Indeed, typical CO-to-optical size ratio is $\sim0.5$ \citep{Young1995}, and in general of the order of unity, as confirmed independently by NOEMA observations of this work (see e.g., Fig.~\ref{fig:allBCGs_optical_CO}). The wobbler-switching mode was used for all observations to minimize the impact of atmospheric variability.
The Wideband Line Multiple Autocorrelator (WILMA) was used to cover the 4$\times$4~GHz bandwidth in each linear polarization.   We also simultaneously recorded the data with the fast Fourier transform spectrometers (FTS) as a backup at 200 kHz resolution.

We encountered good weather conditions during our observations. Overall, we had average system temperatures typical of the winter season and equal to $T_{\rm sys}\simeq$120, 200, 280~K, at 3.6~mm, 1.8~mm, and 1.2~mm, respectively. 
The data reduction and analysis were performed using the {\tt\string CLASS} software of the {\tt\string GILDAS}  package.\footnote{https://www.iram.fr/IRAMFR/GILDAS/}

We applied standard efficiency corrections to convert i) the antenna temperature Ta$^\ast$ into the main beam temperature $T_{\rm mb}$, and then ii) $T_{\rm mb}$ into the corresponding CO line flux, with a conversion factor of 5~Jy/K. In particular, we adopted the following efficiency  corrections: {$T_{\rm mb}/T{\rm a}^\ast$} = 1.17, 1.27, and 1.56 at 3.6~mm, 1.8~mm, and 1.2~mm, respectively.\footnote{https://www.iram.es/IRAMES/mainWiki/Iram30mEfficiencies}.

We clearly detected the BCG at both CO(2$\rightarrow$1) and CO(3$\rightarrow$2) transitions with a signal-to-noise ratio ${\rm S/N}\gtrsim8$. Figure~\ref{fig:COspectra_IRAM30m_MACS1447} displays the corresponding spectra. We fit the total emission using a single Gaussian for each of the CO(2$\rightarrow$1) and CO(3$\rightarrow$2) spectra, after removing the baseline with a linear fit.

CO(1$\rightarrow$0) was not detected, and we reached a root mean square (rms) of 3~mJy for velocity channels of 71~km~s$^{-1}$ after removing the baseline with a linear fit. This yielded a 3$\sigma$ upper limit for the velocity integrated flux of $S_{\rm CO(J\rightarrow J-1)}\,\Delta\varv<1.2$~Jy~km~s$^{-1}$, where we assumed a velocity resolution and width of 250~km~s$^{-1}$, which is typical of massive ellipticals such as MACS~1447 \citep[e.g.,][]{Edge2001,Dunne2021}. The value we adopted is 
the average between those found with the IRAM~30m for the full width at half maximum (FWHM) of the CO(2$\rightarrow$1) and CO(3$\rightarrow$2) transitions of MACS~1447.

\begin{table*}[tb]
\caption{Results of our molecular gas observations with NOEMA.}
\centering
\begin{adjustwidth}{-0cm}{}
\begin{center}
\begin{tabular}{cccccccccccccc}
\hline
\hline
   Galaxy  &  Transition  &  Component &  Flux  & Velocity & FWHM &  $L^\prime_{\rm line}$  \\
   &  &  & (Jy~km~s$^{-1}$)  &  (km~s$^{-1}$) & (km~s$^{-1}$)  & (K~km~s$^{-1}$~pc$^{2}$) \\
 (1) & (2) & (3) & (4) & (5)  & (6) & (7) \\
 \hline
RX~1532 & CO(3$\rightarrow$2) & Total &   $22.73\pm0.30$  &    $-52\pm2$  &  $349\pm6$ & $(1.68\pm0.02)\times10^{10}$\\\\
 & & 0 &  $9.80\pm0.09$ &   $-78\pm2$ &  $353\pm4$ & $(7.26\pm0.07)\times10^9$\\
  & & E &  $3.00\pm0.08$ &   $-41\pm4$ &  $332\pm11$ & $(2.22\pm0.06)\times10^9$\\
  & & W &  $3.90\pm0.09$ &   $-21\pm4$ &  $344\pm10$ & $(2.89\pm0.07)\times10^9$\\
  & & S &  $2.20\pm0.11$ &   $-88\pm8$ &  $311\pm19$ & $(1.63\pm0.08)\times10^9$\\
  \hline
MACS~1447 & CO(3$\rightarrow$2) & Total &  $4.06\pm0.60$ &    $-119\pm6$  &    $130\pm8$ &  $(6.58\pm0.72)\times10^9$\\
 & & &   $4.11\pm0.67$  &    $34\pm14$   &   $197\pm29$  &   \\\\
 & & 0 &  $4.40\pm0.06$ &   $-48\pm2$ &  $241\pm3$ & $(3.54\pm0.05)\times10^9$\\
  & & E &  $1.85\pm0.05$ &   $86\pm3$ &  $257\pm8$ & $(1.49\pm0.04)\times10^9$\\
  & & W &  $1.52\pm0.04$ &   $-127\pm2$ &  $137\pm4$ & $(1.22\pm0.03)\times10^9$\\
   \hline
CHIPS~1911 & CO(2$\rightarrow$1) & Total &  $3.94\pm0.40$ &   $-181\pm2$  &    $175\pm10$  & $(3.46\pm0.18)\times10^{10}$\\
 &  &  & $7.33\pm0.41$  &   $-79\pm10$  &    $448\pm14$ \\\\
 & & 0 & $9.47\pm0.16$ &     $-158\pm2$ &  $295\pm7$ & $(2.91\pm0.05)\times10^{10}$\\
  & & N & $0.40\pm0.04$ &     $-38\pm7$ &  $140\pm18$ & $(1.23\pm0.12)\times10^9$\\  
\hline
\end{tabular}
\end{center}
\tablefoot{Column description: (1) galaxy name;  (2) molecular gas transition; (3) observed component, where "Total" stands for the total flux integrated over the full extension of the source, while 0, N, S, W, and E denote that the flux has been measured over the central, northen, southern, western, and eastern sub-regions, respectively, as illustrated in Fig.~\ref{fig:NOEMAmaps}; (4) velocity integrated flux of the line; (5) relative velocity with respect to that corresponding to the source redshift; (6) line full width at half maximum; (7) velocity integrated line luminosity. For a given BCG and component, multiple entries for the columns (4-6) refer to Gaussian fits where two curves have been used to fit each spectrum, while one Gaussian is used instead as default.}
\label{tab:properties_mol_gas}
\end{adjustwidth}
\end{table*}

\subsection{NOEMA observations} \label{sec:NOEMAobservations}
We observed the three BCGs with the NOEMA interferometer situated at the Plateau de Bure in the French Alps. We used the PolyFix correlator, which covers a total bandwidth of 15.5~GHz in each linear polarization, split between the lower and upper side bands.
We set the phase center of the observations equal to coordinates of the BCGs, as given in Table~\ref{tab:BCG_properties}. 

\subsection{RX~1532}\label{sec:NOEMAobservations_RX1532}
RX~1532 ($z=0.361$) was observed observed in C-configuration, as part of the program  S23BY (PI:~Castignani). The source was observed with 10 antennas on November 8 and 29, 2023, and with 12 antennas on December 19, 2023. We primarily targeted the CO(3$\rightarrow$2) transition, redshifted at $\sim1.2$~mm, given the redshift of the source. The total on-source observing time was 5.53~hr.  We adopted a tuning frequency  in the upper side band, corresponding to the redshifted CO(3$\rightarrow$2) observed frequency. 
The baseline range was 24--368\,m. The program was executed under fair weather conditions, with a system temperature between $T_{\rm sys}=100-200$\,K on November 8 and 29, 2023, and   
between $T_{\rm sys}=200-300$\,K on December 19, 2023. The precipitable water vapor column was between $\sim1-2$\,mm. 
The sources J1514+197, J1611+343, and J1600+335 were used as phase and amplitude calibrators, while   3C~273, MWC~349, 3C~84, and LKHA~101  served as bandwidth and flux calibrators.
The half power primary beam width is $19.9''$, at the tuning frequency of 253.851~GHz.
We adopted natural weighting, yielding a synthesized beam of 
$0.8^{\prime\prime}\times0.6^{\prime\prime}$ (PA=19$^\circ$). We then re-binned the spectral axis at a resolution of 20.3~MHz (i.e., 24~km~s$^{-1}$).  The resulting rms is 0.6\,mJy\,beam$^{-1}$.

\subsection{MACS~1447}\label{sec:NOEMAobservations_MACS1447}
MACS~1447 ($z=0.376$) was observed on 22 December 2021 in C-configuration using 10 antennas, as part of the program
W21CL (PI:~Salomé). We primarily targeted the CO(3$\rightarrow$2) transition, redshifted at $\sim1.2$~mm, given the redshift of the source. 
We adopted a tuning frequency  in the upper side band, corresponding to the redshifted CO(3$\rightarrow$2) observed frequency. The total on-source integration time was 5~h.
The baseline range was 24--400\,m. The program was executed under good weather conditions,
with seeing equal to $0.23^{\prime\prime}$, and with a system temperature $T_{\rm sys}=90$\,K. The precipitable water vapor column was $0.6$\,mm.
The sources 3C273, 2010+723 and 1502+106 were used as phase and amplitude calibrators, while  MWC349  served as bandwidth and flux calibrators.
The half power primary beam width is $20.1''$, at the tuning frequency of 251.305~GHz.
We adopted natural weighting, yielding a synthesized beam of 
$1.36^{\prime\prime}\times0.52^{\prime\prime}$ (PA=11$^\circ$). We then re-binned the spectral axis at a resolution of 20.1~MHz (i.e., 24~km~s$^{-1}$).  The resulting rms is 0.2\,mJy\,beam$^{-1}$.

\subsection{CHIPS~1911}\label{sec:NOEMAobservations_CHIPS1911}
CHIPS~1911 ($z=0.485$) was observed on 18 July and 27 August 2021 in C-configuration using 10 antennas, as part of the program
S21BX (PI:~Combes). We primarily targeted the CO(2$\rightarrow$1) transition, redshifted at $\sim1.9$~mm, given the redshift of the source.
We adopted a tuning frequency  in the upper side band, corresponding to the redshifted CO(2$\rightarrow$1) observed frequency. The total on-source integration time was 3.4~h in July and 3.8~h in August. The baseline range was 32--175\,m. The program was executed under correct and good weather conditions,
with seeing between $0.8''$ and $1''$, and with a system temperature between $T_{\rm sys}=150-250$\,K in July 2021, and
between $T_{\rm sys}=80-110$\,K in August 2021. The precipitable water vapor column was between $\sim5-8$\,mm,
and $\sim1-3$\,mm, respectively.
The sources 3C454.3, 2010+723, 1851+488 and 1828+487 were used as phase and amplitude calibrators, while  MWC349  served as bandwidth and flux calibrators. The half power primary beam width is $32.5''$, at the tuning frequency of 155.244~GHz. We adopted natural weighting, yielding a synthesized beam of 
$2.7^{\prime\prime}\times2.3^{\prime\prime}$ (PA=71$^\circ$). We then re-binned the spectral axis at a resolution of 25.9~MHz, that is, 50~km~s$^{-1}$.  The resulting rms is 0.28\,mJy\,beam$^{-1}$.\\

For all three BCGs, we reduced NOEMA data using the \textsf{clic} package of the \textsf{GILDAS} software to obtain the final ($u$,$v$) tables.
We then imaged the visibilities using the \textsf{MAPPING} software  of \textsf{GILDAS}.

\section{Results}
\label{sec:results}
\subsection{CO emission}\label{sec:COemission}

For all three BCGs of this work our NOEMA observations clearly detected extended emission in CO(3$\rightarrow$2) for RX~1532 and MACS~1447 and CO(2$\rightarrow$1) for CHIPS~1911, up to $\sim5$~arcsec from the phase center, that is, the BCG projected coordinates. Additional lines could have been serendipitously detected from companion cluster members, as our observations could potentially detect them within the primary beam, that is, within a radius of 50.0, 52.3, 97.5 kpc from the BCG, respectively. However, when inspecting the entire datacube, no other emission line was detected in the upper side band nor in the lower side band.

Figure~\ref{fig:NOEMAmaps} shows the maps of CO maxima, as well as moment 0, 1, and 2 CO maps for all three BCGs, which are displayed after subtracting the continuum emission (see Sect.~\ref{sec:NOEMAcontinuum}). The intensity maps and the maps of maxima are displayed in the top two rows of the figure. Several connected CO emitting components are identified in the maps. These are highlighted with dashed lines in the intensity maps, which delimit the sub-regions denoted as 0, N, S, W, and E. Namely we identified a central CO emitting component (0) coincident with RX~1532 host galaxy, as well as nearby eastern (E), western (W), and southern (S) emitting regions. Similarly, for MACS~1447 we identified a central component (0) and nearby eastern (E) and western (W) sub-regions. For CHIPS~1911, a central CO emitting component (0), co-spatial with the host galaxy, and a northern (N) extension are clearly identified from the intensity and maxima maps. For each BCG, the region delimited by the unions of these sub-regions, hereafter denoted as "Total," defines the overall emitting area in the projected space.

By integrating over these "Total" regions, we extracted CO spectra for all three BCGs, separately, and we report them in Fig.~\ref{fig:COspectra_total}.
In Figs~\ref{fig:RX1532_COspectra_regions}, \ref{fig:MACS1447_COspectra_regions}, and \ref{fig:CHIPS1911_COspectra_regions} of the   Appendix~\ref{sec:NOEMAspectra_subregions}, we report instead the CO spectra associated with all sub-regions of the three BCGs, as defined from  the intensity maps. All spectra show that CO emission is clearly detected at high S/N for all BCGs and all their sub-regions, well fit by single or double Gaussian components.

Our analysis shows a complex molecular gas morphology for each BCG {(see e.g., moment-0 maps of Fig.~\ref{fig:NOEMAmaps}),} where the bulk of the molecular gas is co-spatial with the host galaxy (i.e., with the central sub-region 0). More specifically, for RX~1532, the component 0 accounts for about $\sim50\%$ of the total CO flux, while each of the nearby E, W, and S sub-regions, which extend up to 10-20~kpc in radius, accounts for $\sim(10-20)\%$ of the total CO 
flux, yielding a relative contribution that is similar to that found in the literature for clumps and filaments surrounding local BCGs \citep{Olivares2019}. Similar results are found for MACS~1447, where the central component 0 comprises about half of the total emission, while the remaining 50\% is equally split between the nearby E and W regions, which span altogether a region of $\sim30$~kpc along the 
east-west direction. For a comparison, using ALMA observations, \citet{Russell2019} found similar results for the star-bursting Phoenix BCG, with $\sim50\%$ of molecular gas found in the filaments and an extension of up to 10-20~kpc. Last, the molecular gas emission of CHIPS~1911 is more concentrated, with $\sim96\%$ of the total flux comprised within the central component (0) and the rest in the northern extension (N).  In Table~\ref{tab:properties_mol_gas} we summarize the results of our analysis of NOEMA data. Namely, for each source and CO emitting component, we report the velocity integrated flux, relative recession velocity with respect to the BCG redshift, FWHM of the line, and its velocity integrated luminosity, as determined in Sect.~\ref{sec:gas_reservoirs}.

\begin{figure}[h!]
\centering
\includegraphics[width=0.4\textwidth]{{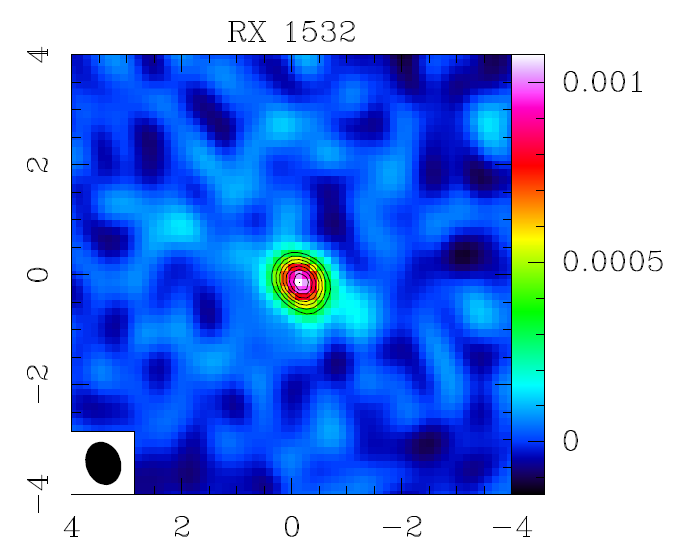}}\vspace{0.5cm}
\includegraphics[width=0.4\textwidth]{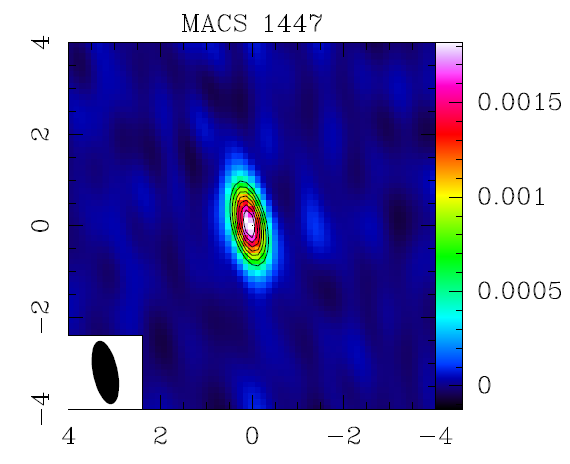}\vspace{0.5cm}
\includegraphics[width=0.4\textwidth]{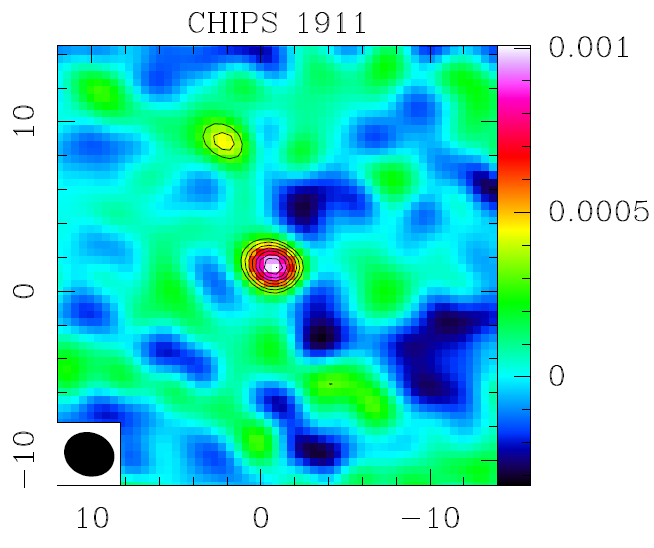}
\caption{From top to bottom: lower side band continuum maps of RX~1532 (top), MACS~1447 (center), and CHIPS~1911 (bottom). Coordinates are reported as separations from the phase center. The color scale shows the continuum flux in units of Jy.}
\label{fig:continuum_BCGs_maps}
\end{figure}

\renewcommand{\arraystretch}{2}
\begin{table}[tb]
\caption{Radio continuum properties.}
\centering
\begin{adjustbox}{width=0.48\textwidth, left}
\begin{tabular}{cccccccccccccc}
\hline
\hline
   Galaxy  &  $\nu$ &  $F_\nu$ & $\alpha$ & $L_{\rm 1.4~GHz}$\\
   & (GHz) &  (mJy) & & {($10^{25}$~W~Hz$^{-1}$)}  \\
 (1) & (2) & (3) & (4) & (5) \\
 \hline
RX~1532 & 240.256 & $1.08\pm0.03$ & $0.60\pm0.01$ & $0.91\pm0.03$  \\
 & 1.4 & $23.3\pm0.8$  \\
\hline
MACS~1447 & 237.756 & $1.82\pm0.01$ & $0.62\pm0.01$ & $1.88\pm0.06$ \\
 & 1.4 & $43.6\pm1.4$ \\
 \hline
CHIPS~1911 & 141.256 & $1.11\pm0.02$ &  $0.62\pm0.01$ & $1.47\pm0.06$\\
 & 1.4 & $19.2\pm0.8$ \\
 \hline
\end{tabular}
\end{adjustbox}
\tablefoot{Column description: (1)  galaxy name; (2) observer frame frequency; (3) continuum radio flux from NOEMA observations presented in this work, evaluated using the lower side band, and from NRAO VLA Sky Survey (NVSS) at 1.4~GHz \citep{Condon1998}; (4) radio spectral index $\alpha$ inferred using NOEMA (at millimeter wavelengths) and 21~cm flux, where a power law $F_\nu\propto\nu^{-\alpha}$ is assumed; (5) rest frame 1.4~GHz  radio luminosity density.}
\label{tab:radio_fluxes}
\end{table}

\begin{table*}[tb]
\caption{Molecular gas properties of the BCGs.} 
\centering
\begin{adjustbox}{width=1.0\textwidth, left}
\begin{tabular}{ccccccccccccc}
\hline\hline
 Galaxy  &  $z_{spec}$ & CO(J$\rightarrow$J-1)  & $\nu_{\rm obs}$ & $S_{\rm CO(J\rightarrow J-1)}\,\Delta\varv$   & FWHM &  $M_{H_2}$ & $\tau_{\rm dep}$ & $\frac{M_{H_2}}{M_\star}$  & $\tau_{\rm dep, MS}$ & $\big(\frac{M_{H_2}}{M_\star}\big)_{\rm MS}$  \\
   &  & & (GHz) & (Jy~km~s$^{-1}$)  & (km~s$^{-1}$) & ($10^{10}~M_\odot$) & ($10^9$~yr) &  & ($10^9$~yr) & \\
 (1) & (2) & (3) & (4) & (5) & (6) & (7) & (8) & (9) & (10) & (11) \\
 \hline
RX~1532 &    0.361 & 1$\rightarrow$0 & 84.696 & $3.0\pm0.4$  & $369\pm51$ & $8.7\pm1.1$ &$1.21\pm0.25$  & $0.65\pm0.08$ & $1.16^{+0.17}_{-0.15}$ & $0.09^{+0.09}_{-0.08}$\\
       &          & 3$\rightarrow$2 & 254.075 & $20.2\pm1.6$ & $390\pm36$  & & & &  & \\
       \hline
MACS~1447 &    0.376 & 1$\rightarrow$0 & 83.773  & $<1.2$ & ---    & $8.9\pm1.1$ & $1.27\pm0.22$   & $0.14\pm0.02$  & $1.32^{+0.28}_{-0.23}$ & $0.06^{+0.05}_{-0.05}$ \\
       &          & 2$\rightarrow$1 & 167.542 &  $9.0\pm1.1$ & $273\pm37$ & & &  & \\ 
       &          & 3$\rightarrow$2 & 251.305 &  $16.4\pm1.9$ & $221\pm36$ & & &  & \\ 
       \hline
       CHIPS~1911 &    0.485 & 2$\rightarrow$1 & 155.244 & $3.94\pm0.40$ & $175\pm10$    & $18.9\pm1.0$ & $1.47\pm0.21$ & $0.86^{+0.14}_{-0.40}$  & $1.15^{+0.20}_{-0.17}$  & $0.11^{+0.16}_{-0.06}$ \\ 
        &    &  &  & $7.33\pm0.41$ & $448\pm14$    &  &  &  &  &  \\
\hline
\end{tabular}
\end{adjustbox}
\tablefoot{For RX~1532 and MACS~1447, molecular gas properties are derived from IRAM~30m observations as in \citet{Castignani2020a} and this work, respectively. For CHIPS~1911, molecular gas properties correspond to the emission observed with NOEMA (component: Total, this work). Column description : (1) galaxy name;  (2) spectroscopic redshift as in Table~\ref{tab:BCG_properties}; (3-4) CO(J$\rightarrow$J-1) transition and observer frame frequency; (5-6) CO(J$\rightarrow$J-1) velocity integrated flux and FWHM, both derived from single Gaussian fit to the CO(J$\rightarrow$J-1) emission line; (7) molecular gas mass obtained with $\alpha_{\rm CO}=4.36~M_\odot\,({\rm K~km~s}^{-1}~{\rm pc}^2)^{-1}$, where the CO(1$\rightarrow$0) flux is used for RX~1532, while for MACS~1447 and CHIPS~1911 the CO(2$\rightarrow$1) fluxes are used instead, assuming an excitation ratio $r_{21}=0.8$; (8) depletion timescale  $\tau_{\rm dep}=M_{H_2}/{\rm SFR}$; (9) molecular gas-to-stellar mass ratio; (10-11) depletion timescale and molecular gas-to-stellar mass ratio predicted for MS field galaxies with redshift and stellar mass of our targets, following \citet{Tacconi2018}.}
\label{tab:BCG_properties_mol_gas}
\end{table*}

\subsection{CO kinematics}\label{sec:COkinematics}
{Previous studies of local BCGs invariably found a complex morphology of their molecular gas reservoirs in terms of filamentary structures and extended clumps \citep{Olivares2019,Russell2019}. Similarly,  the kinematics of the molecular gas reservoirs for the cool-core BCGs of this work appears complex.} For each BCG, the CO components have moderate relative velocity displacements with respect to the BCG redshift,   (see Table~\ref{tab:properties_mol_gas}). Altogether, the relative velocity displacements are more pronounced in the case of MACS~1447 and CHIPS~1911, resulting in asymmetric velocity spectra  (Fig.~\ref{fig:COspectra_total}), whose fits required modeling with multiple Gaussian components.

Furthermore, moment-1 maps show that the BCG molecular gas is not rotationally supported. 
Overall, we do not find rotating molecular gas disks in the BCGs, suggesting dynamically perturbed molecular gas reservoirs. Altogether, the components 0, N, S, E, and W of the BCGs have small recession velocities $\sim100$~km~s$^{-1}$, relative to the barycenter of the systems. Namely, these values are smaller than the BCG stellar velocity dispersion, which we estimated to be equal to 215, 327, and 346~km~s$^{-1}$ for RX~1532, MACS~1447, and CHIPS~1911, respectively. For these estimates we used the fiducial scaling relation by \citet{Zahid2016} between the stellar mass and the stellar velocity dispersion. The observed molecular gas recession velocities for the different BCG components are also much lower
than the escape velocity for the BCGs, of the order of $\sim1\,000$~km~s$^{-1}$ \citep[e.g.,][]{Russell2019}.  Therefore, all molecular gas components are  gravitationally bound to the galaxies. These findings are in agreement with those found in previous studies of nearby BCGs 
\citep{McNamara2014,Russell2016,Russell2019,Olivares2019}.

Consistently with this picture, a globally coherent distribution of the molecular gas reservoirs emerges also from the moment-2 maps of the BCGs (Fig.~\ref{fig:continuum_BCGs_maps}). The velocity dispersion increases from the outer regions up to $\sim(300-400)$~km~s$^{-1}$ at the BCG center, inside the central component 0. This is reasonable, as the gravitational potential is deeper closer to the BCG (and cluster) center.
However, when comparing the different components, the velocity widths are fairly similar, in the range between $\sim(140-350)$~km~s$^{-1}$ (Table~\ref{tab:properties_mol_gas}).

Interestingly, our observations also tend to disfavor the scenario where the molecular gas components (0, N, S, E, W) surrounding the BCGs are gas blobs that are free-falling to the bottom of the deep gravitational potential of the cluster. Indeed, previous  studies have found that in the case of free-falling, gas blobs should be accelerated up to velocities of a few hundred km~s$^{-1}$ over distances of a
few kiloparsecs \citep{Lim2008,Russell2016,Vantyghem2016}. 
While the molecular gas distribution and its velocity field suggest a planar distribution of the gas reservoirs where small recession velocities may not be surprising, such predicted high velocities are not observed in this work, nor in nearby BCGs \citep[e.g.,][]{Olivares2019,Russell2019}.

\subsection{Continuum emission}\label{sec:NOEMAcontinuum}
In addition to the molecular gas, we detected continuum emission in both the lower and upper side bands co-spatial with the BCGs. Figure~\ref{fig:continuum_BCGs_maps} displays the continuum maps in the lower side band. For all BCGs, the continuum emission is point-like, detected with similar flux densities in the range $\sim(1.1-1.8)$~mJy. We also looked for low frequency radio emission from the NRAO VLA Sky Survey (NVSS) at 1.4~GHz \citep{Condon1998}. All BCGs have clear 1.4~GHz NVSS counterparts, with flux densities in the range $\sim(19.2-43.6)$~mJy. By combining continuum emission at millimeter and 21~cm wavelengths from NOEMA and NVSS, respectively, we inferred a spectral index $\alpha\simeq0.6$, for all BCGs, where we assumed a power law relation between the flux density ($F_\nu$) and the observed frequency ($\nu$), that is, $F_\nu\propto\nu^{-\alpha}$.

The point-like morphology and the observed steep radio spectrum suggest that the observed millimeter continuum emission is mostly due to synchrotron emission from the AGN. 
{However, it is worth mentioning that optically thin synchrotron emission could be steeper than estimated. Indeed, while the 1.4~GHz radio emission is likely due to the radio lobes and jets, the continuum at frequencies $\gtrsim100$~GHz, probed by NOEMA observations, is largely dominated by the flat-spectrum core of the AGN \citep[e.g.,][]{Hogan2015b}.}

Furthermore, by using the 1.4~GHz NVSS fluxes $F_{\rm 1.4~GHz}$ we estimated the isotropic rest frame 1.4~GHz radio luminosity densities as $L_{\rm 1.4~GHz} = 4\pi~F_{\rm 1.4~GHz}~D_L(z)^2~(1+z)^{\alpha-1}$, where $D_L$ is the luminosity distance, evaluated at the BCG redshift $z$.
This procedure yields similar  $L_{\rm 1.4~GHz}$ values for the three BCGs in the range {$\simeq(0.9-1.9)\times10^{25}$~W~Hz$^{-1}$. The three BCGs thus tend to populate the bright end of the BCG radio luminosity function \citep[e.g.,][]{Hogan2015a}. Namely, the three BCGs have radio luminosities typical of bright} FR~I radio galaxies, similar to famous local BCGs such as M87 or Perseus~A, given the classical FR~I~/~FR~II radio power divide 
\citep[$L_{\rm 1.4~GHz}\simeq2\times10^{25}$~{W~Hz}$^{-1}$;][]{Fanaroff_Riley1974}.
In Table~\ref{tab:radio_fluxes} we summarize the radio continuum properties of the BCGs.

\begin{figure*}[t!]
\centering
\includegraphics[width=0.35\textwidth]{{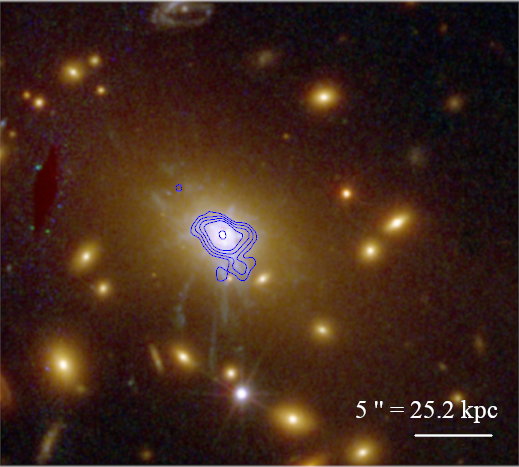}}
\includegraphics[width=0.315\textwidth]{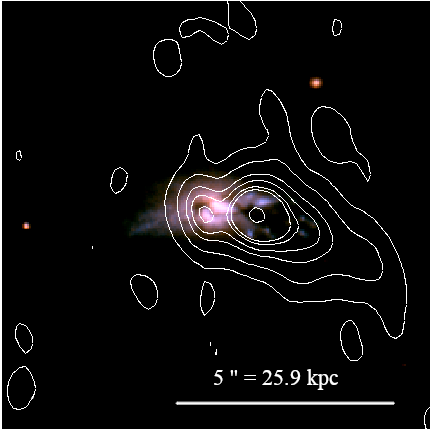}
\includegraphics[width=0.31\textwidth]{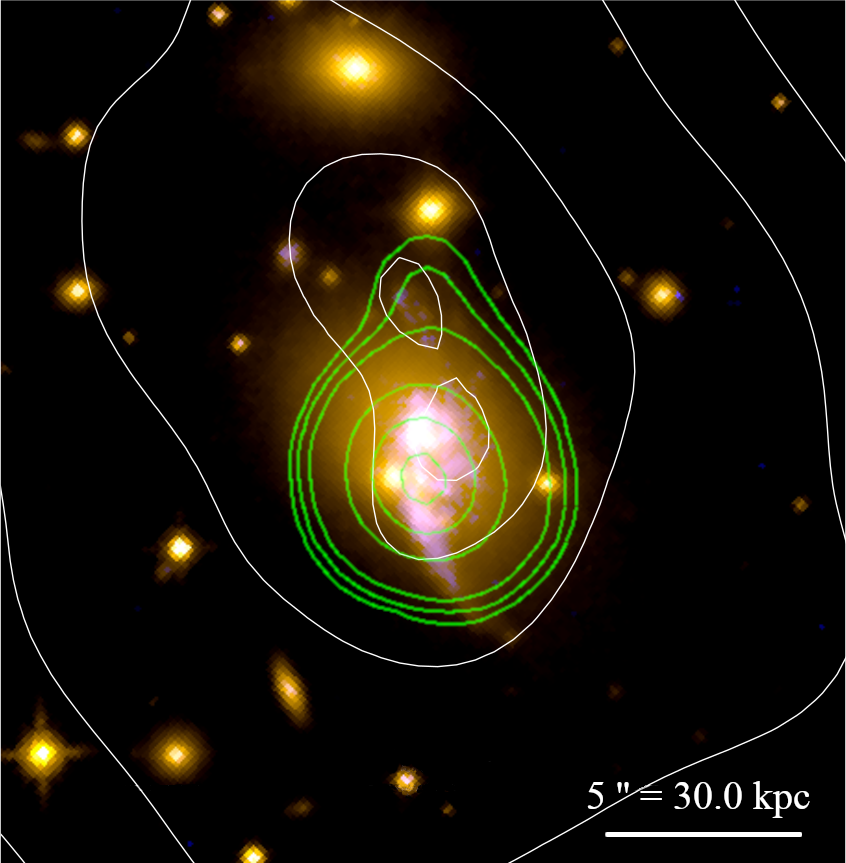}
\caption{From left to right: Color-composite images of RX~1532 (left),  MACS~1447 (center), and  CHIPS~1911 (right). {The RGB image of RX~1532 is taken from \citet{Fogarty2015} and made using {\it HST} WFC3-IR filters F105W + F110W + F125W + F140W + F160W in red, the ACS filters F606W + F625W + F775W + F814W + F850LP in green, and the ACS filters F435W + F475W in blue. The RGB image of MACS~1447 is from \citet[][]{PrasowEmond2020}, and made using the {\it HST} WFC filters F814W (red and blue channels) and F606W (green). For CHIPS~1911, the color-composite image is taken from \citet[][]{Somboonpanyakul2021a} and made using F550M data from the ACS and F110W data from the WFC infrared channel (WFC3-IR).}
CO contours from NOEMA observations reported in this work are superimposed with blue {(left)}, white {(center)}, and green {(right)} solid lines,  respectively. 
{For all three BCGs, contour levels start at S/N~=~3, where the rms noise levels, estimated directly from the CO intensity maps, are equal to 20, 8, and 36~mJy~beam$^{-1}$~km~s$^{-1}$ for RX~1532, MACS~1447, and CHIPS~1911, respectively. Namely, contour levels correspond to (0.06, 0.2, 0.5, 1.0, 5.0)~Jy~beam$^{-1}$~km~s$^{-1}$ in CO(3$\rightarrow$2) for RX~1532; (0.024, 0.2, 0.5, 0.7, 0.9, 1.0, 1.5)~Jy~beam$^{-1}$~km~s$^{-1}$ in CO(3$\rightarrow$2) for MACS~1447; and (0.108, 0.2, 0.4, 1.9, 3.7, 5.5)~Jy~beam$^{-1}$~km~s$^{-1}$ in CO(2$\rightarrow$1) for CHIPS~1911.} For CHIPS~1911, white contours refer to the {\it Chandra} 0.5–7.0 keV emission and are from \citet[][]{Somboonpanyakul2021a}. North is up, east is to the left.} 
\label{fig:allBCGs_optical_CO}
\end{figure*}

\subsection{Molecular gas reservoirs}\label{sec:gas_reservoirs}

For all BCGs, we used the velocity-integrated CO(J$\rightarrow$J-1) fluxes $S_{\rm CO(J\rightarrow J-1)}\,\Delta\varv\ $, in units {of} Jy~km~s$^{-1}$, which were obtained by Gaussian fits of the NOEMA spectra, as described Sect.~\ref{sec:COemission}, to infer velocity integrated CO(J$\rightarrow$J-1) luminosities $L^{\prime}_{\rm CO(J\rightarrow J-1)}$, in units {of} K~km~s$^{-1}$~pc$^2$. 
To this aim, we used Eq.~(3) from \citet{Solomon_VandenBout2005}:
\begin{equation}
\label{eq:LpCO}
 L^{\prime}_{\rm CO(J\rightarrow J-1)}=3.25\times10^7\,S_{\rm CO(J\rightarrow J-1)}\,\Delta\varv\,\nu_{\rm obs}^{-2}\,D_L^2\,(1+z)^{-3}\,,
\end{equation}
where $\nu_{\rm obs}$ is the observer frame frequency in GHz of the CO(J$\rightarrow$J-1) transition, $D_L$ is the luminosity distance in Mpc, and $z$ is the redshift of the BCG. 
Similarly, we used the same procedure to derive $L^{\prime}_{\rm CO(J\rightarrow J-1)}$ luminosities from the fluxes resulting from IRAM~30m observations of MACS~1447 (see Sect.~\ref{sec:IRAM30m_observations}).

We then inferred the total $H_2$ gas masses $M_{H_2}$ of the BCGs by using the velocity-integrated CO line luminosities, that is, 
$M_{H_2} = \alpha_{\rm CO}\;L^{\prime}_{\rm CO(1\rightarrow0)} = \alpha_{\rm CO}\;L^{\prime}_{\rm CO(J\rightarrow J-1)}/r_{\rm J1}$, where $\alpha_{\rm CO}$ is the CO-to-$H_2$ conversion factor and $r_{\rm J1}=L^{\prime}_{\rm CO(J\rightarrow J-1)}/L^{\prime}_{\rm CO(1\rightarrow0)}$ denotes the excitation ratio.
As highlighted in Sect.~\ref{sec:Mstar_SFR}, the three BCGs are star forming. Namely, MACS~1447 and CHIPS~1911 have ${\rm SFR}\lesssim~3\times{\rm SFR}_{\rm MS}$, thus within the characteristic scatter of the MS. RX~1532 has a higher SFR, when normalized to the MS, that is ${\rm SFR}\simeq~(6-8)\times{\rm SFR}_{\rm MS}$. To estimate $H_2$ gas masses, we  assumed a Galactic CO-to-$H_2$ conversion factor of $\alpha_{\rm CO}=4.36~M_\odot\,({\rm K~km~s}^{-1}~{\rm pc}^2)^{-1}$, which is commonly used for star-forming galaxies at the MS.

\citet{Castignani2020a} observed RX~1532 with the IRAM~30m and clearly detected both CO(1$\rightarrow$0) and CO(3$\rightarrow$2) emission, with fluxes of $(3.0\pm0.4)$ and $(20.2\pm1.6)$~Jy~km~s$^{-1}$, respectively. Interestingly, our NOEMA observations yielded a similar CO(3$\rightarrow$2) flux of $(22.73\pm0.30)$~Jy~km~s$^{-1}$. The FWHM values in CO(3$\rightarrow$2) are also similar and equal to $(349\pm6)$~km~s$^{-1}$ (NOEMA) and $(390\pm36)$~km~s$^{-1}$ (IRAM~30m). This agreement strengthens not only the validity of both NOEMA and IRAM~30m analyses, but it also implies that NOEMA observations do not miss any clear extended emission that might have been filtered out by the interferometric observations. Altogether, similarly to \citet{Castignani2020a}, to infer the total $H_2$ gas mass of RX~1532, we used the CO(1$\rightarrow$0) flux obtained with the IRAM~30m observations, which we preferred to the CO(3$\rightarrow$2) ones from NOEMA or the IRAM~30m, as the former allowed us to derive $M_{H_2}$ directly, that is, without the use of the excitation ratio.

Concerning MACS~1447, we performed both IRAM~30m and NOEMA observations in several CO lines, as described earlier. More precisely, thanks to IRAM~30m observations, we were able to derive an upper limit to the CO(1$\rightarrow$0) emission, while both CO(2$\rightarrow$1) and CO(3$\rightarrow$2) were clearly detected. Similarly, thanks to NOEMA observations, we clearly detected CO(3$\rightarrow$2) emission from the BCG. As illustrated in Table~\ref{tab:properties_mol_gas} the total CO(3$\rightarrow$2) NOEMA flux is $(8.17\pm0.89)$~Jy~km~s$^{-1}$, which is a factor of $\sim2$ lower than that found in CO(3$\rightarrow$2) with the IRAM~30m, that is, $(16.4\pm1.9)$~Jy~km~s$^{-1}$. It is thus possible that some extended emission has been filtered out by NOEMA interferometer. Altogether, to estimate $M_{H_2}$ for MACS~1447, we used the CO(2$\rightarrow$1) flux obtained from the IRAM~30m observations, assuming an excitation ratio $r_{21}=0.8$ \citep{Bothwell2013,Birkin2021,denBrok2021,Leroy2022}.  We preferred to use the CO(2$\rightarrow$1) flux to those we obtained for the higher-J CO(3$\rightarrow$2) transition as the latter usually requires higher gas densities and temperatures. Altogether, CO(2$\rightarrow$1) is a good proxy for the total $H_2$ cold gas mass, as an alternative to the CO(1$\rightarrow$0) emission \citep[e.g.,][]{Bolatto2013,Tacconi2018}, for which we have only an upper limit from the IRAM~30m.

Concerning CHIPS~1911, to estimate $M_{H_2}$, we used the total CO(2$\rightarrow$1) flux that  we obtained with our NOEMA observations. These are the only observations in CO that are available for this source.
Altogether, all three BCGs exhibit high $H_2$ gas masses $\simeq9\times10^{10}~M_\odot$ for both RX~1532 and MACS~1447, and $\simeq2\times10^{11}~M_\odot$ for CHIPS~1911. Similarly, the three BCGs have high $H_2$ gas mass to stellar mass ratios, ranging between $\sim(0.14-0.86)$. They also show large CO line widths, between $\sim(200-400)$~km~s$^{-1}$, typical of massive ellipticals. The three BCGs are therefore gas rich, as similarly found in previous studies of distant star forming BCGs  \citep{McDonald2014,Dunne2021,Castignani2022b,Castignani2023}.

\begin{figure*}[h!]
\hspace{1cm}\includegraphics[width=0.25\textwidth]{{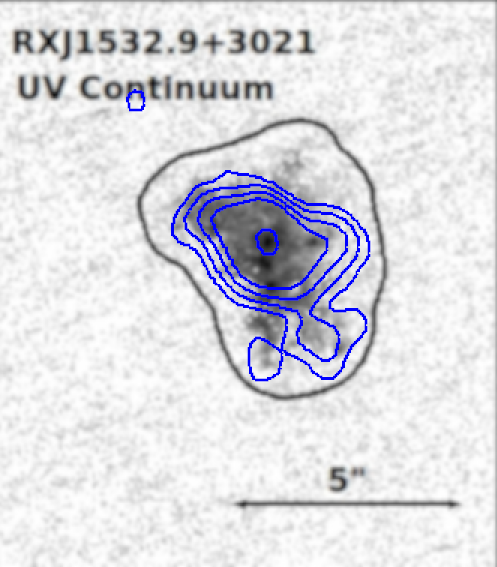}}\hspace{0.2cm}
\includegraphics[width=0.25\textwidth]{{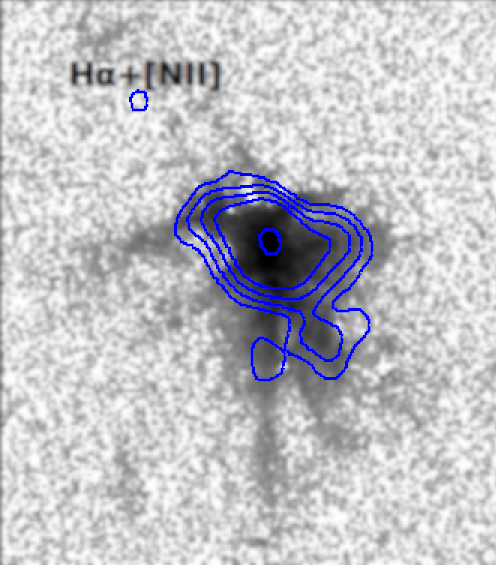}}
\includegraphics[width=0.355\textwidth]{{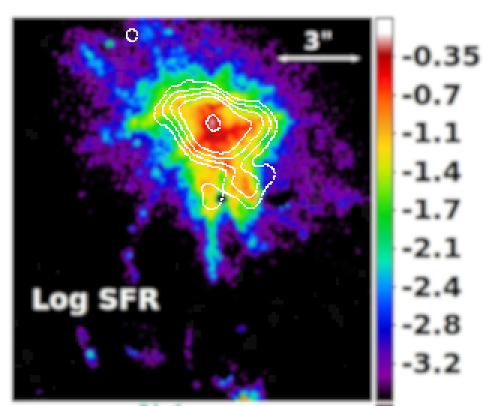}}
\caption{Maps of RX~1532 BCG. Left: Ultraviolet continuum map. The solid curve denotes the region used to estimate the UV luminosity by \citet{Fogarty2015}. Center: H$\alpha$+[NII] by \citet{Fogarty2015}. Right: Star formation rate map reconstructed by \citet{Fogarty2015}. The color bar is in logarithmic units of $\log[{\rm SFR}/(M_\odot {\rm yr}^{-1})]$. Blue (left and center) and white contours (right panel) correspond to the CO(3$\rightarrow2$) contour levels from NOEMA, as reported in Fig.~\ref{fig:allBCGs_optical_CO}. North is up, east is to the left.}
\label{fig:RX1532_SF_CO_maps}
\end{figure*}

Furthermore, we used the $H_2$ gas mass to estimate the depletion time $\tau_{\rm dep}=M_{H_2}/{\rm SFR}$, which is the characteristic timescale at which the molecular gas reservoir is depleted. To compute $\tau_{\rm dep}$ , we used the SFR estimates reported in Table~\ref{tab:BCG_properties}. In case of multiple SFR estimates, we averaged them before computing the depletion time. 

Finally, as a comparison, we computed the depletion time $\tau_{\rm dep, MS}$ and the molecular gas-to-stellar mass ratio $\big(\frac{M_{H_2}}{M_\star}\big)_{\rm MS}$ for MS galaxies in the field with a redshift and stellar mass equal to those of the BCGs, by using the empirical prescriptions by \citet{Tacconi2018}, calibrated to the CO-to-$H_2$ conversion factor of $\alpha_{\rm CO}=4.36~M_\odot\,({\rm K~km~s}^{-1}~{\rm pc}^2)^{-1}$ we used here. 
In Table~\ref{tab:BCG_properties_mol_gas} we summarize the global molecular gas properties of the three BCGs.

Altogether, the above comparison confirms that the three BCGs are gas rich, having $M_{H_2}/M_\star$ ratios that are a factor $\sim(2-8)$ larger than the MS value of field galaxies. However, such large molecular gas reservoirs appear to be depleted with a normal time scale $\tau_{\rm dep}\simeq(1.2-1.5)$~Gyr, that is, similar to that of MS galaxies, at fixed stellar mass and redshift. These results tend to favor  the scenario where strong environmental processes  occur in the BCGs, as we further discuss in Sect.~\ref{sec:discussion}.

\section{Discussion}
\label{sec:discussion}

As outlined in the previous section, on the basis of the global stellar, star formation, and gas properties of the three BCGs, effective mechanisms of gas processing are likely needed to explain simultaneously the high $M_{H_2}/M_\star$ ratios, the high SFRs, and the normal depletion times. We thus propose that relatively recent gas infall or condensation should have occurred, via cooling of the ICM (i.e., flow of cooling gas), via compression of the gas ultimately feeding the star formation, or via mergers with cluster core companions. In fact, these processes are able to effectively replenish cold gas reservoirs of the BCGs, while at the same time sustaining their ongoing star formation with the a normal (main sequence) consumption time scale. As further discussed below, our NOEMA observations likely witness such environmental processes at play.

In Fig~\ref{fig:allBCGs_optical_CO} we display the color-composite images of the three BCGs, using publicly available data from the {\it Hubble Space Telescope (HST)}. {The images reveal the stellar emission of the BCGs, as further discussed below.} 
We superimpose to them the contour levels in CO(3$\rightarrow$2) for RX~1532 and MACS~1447, and in CO(2$\rightarrow$1) for CHIPS~1911, from the NOEMA observations of the present work. {We refer to the caption of Fig.~\ref{fig:allBCGs_optical_CO} for further details.}

\subsection{RX~1532: Cooling filaments}\label{sec:RX1532_discussion}
Visual inspection of the color-composite image of RX~1532 shown in Fig.~\ref{fig:allBCGs_optical_CO} reveals filamentary structures extended towards all directions surrounding the BCG core.
{In addition, we independently} retrieved publicly available {\it HST} images of RX~1532 from the Hubble Legacy Archive (HLA)\footnote{https://hla.stsci.edu/}. We found that the star forming core of the BCG and well contrasted filaments surrounding the BCG at all directions are clearly visible in the F475W and F850LP {\it HST} filters, {which are used for the RGB image in Fig.~\ref{fig:allBCGs_optical_CO}. The two filters are} centered around the redshifted [OII] and H$\alpha$ emission lines of the BCG, respectively, and thus probe ionized gas tracing ongoing star formation. Similar morphology is displayed in Fig.~\ref{fig:RX1532_SF_CO_maps}, where UV luminosity, H$\alpha$+[NII], and SFR maps of RX~1532 by \citet{Fogarty2015} are shown.

The CO(3$\rightarrow$2) emission well recovers the complex morphology of the BCG, both in its stellar and ionized gas content. In particular, the peak of the CO emission is co-spatial with the core of the BCG (Figs.~\ref{fig:allBCGs_optical_CO}, \ref{fig:RX1532_SF_CO_maps}), while we also detected two extended CO components, to the east and to the west of the BCG core, that we denoted as E and W, respectively (see Fig.~\ref{fig:NOEMAmaps}). Interestingly, these two were previously identified as star forming elongated structures by means of stellar analysis performed by \citet[][see their Fig.~15]{Fogarty2015}, which we now detect in molecular gas. Similarly, two southern filaments were identified by \citet{Fogarty2015}, both in ionized gas and in UV continuum (see Fig.~\ref{fig:RX1532_SF_CO_maps}). 
These southern filaments are thus star forming, as appreciated by visual inspection of the SFR density map (Fig.~\ref{fig:RX1532_SF_CO_maps}, right). The most star-forming  filament among the two, at south-west, is co-spatial with a filamentary structure of cold molecular gas (component S in Fig.~~\ref{fig:NOEMAmaps}), extending a few kpc in length to the south of the BCG. 

Furthermore, velocities of cold molecular gas are $\simeq100$~km~s$^{-1}$ (Fig.~\ref{fig:NOEMAmaps}), thus much lower than those of a few hundred km~s$^{-1}$, that are predicted in case of free-falling gas (see Sect.~\ref{sec:COkinematics}, and references therein). They  are instead in agreement with the kinematics found for nearby BCGs \citep{Olivares2019} and with the velocities predicted by simulations of slowly infalling cold gas \citep{Li_Bryan2014, Gaspari2018}.  Indeed, the heating due to the AGN is able to produce turbulence, triggering cold gas condensation, and precipitation  \citep{Voit2015,Voit2018,Gaspari2015,Gaspari2017}. Simulations show that thermal instabilities lead the gas to cool as soon as the cooling time ($t_{\rm cool}$) over the free-fall time ($t_{\rm ff}$) satisfies the relation $t_{\rm cool}/t_{\rm ff}\lesssim10$ 
\citep[e.g.,][]{Sharma2012}. Our analysis thus supports the scenario that RX~1532 is surrounded by cooling filaments of infalling gas, observed in both the ionized and cold gas phases. In terms of complex morphology, kinematics, and kiloparsec-scale length, these filaments of gas resemble the filamentary structures and clumps of gas that have been found in previous work \citep[e.g.,][]{Olivares2019,Russell2019,Fogarty2019} in the surroundings of  nearby and intermediate-$z$ cool-core BCGs such as Centaurus ($z=0.010$), RX~1539.5-8335 ($z=0.076$), Abell~S1101 ($z=0.056$), and  MACS~1932 ($z= 0.352$).

{Interestingly, \citet{Hlavacek-Larrondo2013} found the presence of X-ray cavities to the north-west and south-east of the BCG, likely inflated by the AGN jets.}
As pointed out earlier by \citet{Castignani2020a}, high star formation activity observed in RX~1532, together with the cool-core cluster environment and its associated low entropy floor, ultimately favor the condensation and the inflow of gas onto the BCG itself.

\subsection{MACS~1447: Ram pressure stripping {or sloshing motions}}\label{sec:MACS1447_discussion}

Figure~\ref{fig:allBCGs_optical_CO} displays the complex optical morphology of MACS~1447. As highlighted by \citet{PrasowEmond2020}, the BCG is surrounded by optically bright dusty filaments up to $\sim25$~kpc in size, with the two most prominent ones extending to south-west and  north-west
directions from the BCG. Similarly to what we have done for RX~1532, we retrieved publicly available {\it HST} images from the HLA archive. Interestingly, we checked that both filamentary structures are nicely captured by the {\it HST} images obtained with the broad-band F606W and F814W filters, independently, {which are used in the RGB image shown in Fig.~\ref{fig:allBCGs_optical_CO} and are} centered around the redshifted [O~II] and H$\alpha$ emission lines, respectively. See also Sect.~2.3 of \citet{PrasowEmond2020}, for further discussion. Visual inspection of these images and Fig.~\ref{fig:allBCGs_optical_CO} with the CO(3$\rightarrow$2) contours overlaid, shows that the molecular gas has a globally good overlap with the stellar BCG morphology, as traced by {\it HST}. Furthermore, the prominent stellar filament extending to the south-west is spatially coincident with the south-west molecular gas component W (see Fig.~\ref{fig:NOEMAmaps}). This implies that the molecular gas is co-spatial with the ionized gas in the filament, and ultimately feeds its star formation. 

The X-ray plume to the west of the BCG \citet{PrasowEmond2020} 
is globally co-spatial with the western filaments seen with {\it HST}, and is located in the proximity of the CO(3$\rightarrow$2) emission line peak of the component 0 (Fig.~\ref{fig:NOEMAmaps}), suggesting that strong condensation is at the base of the south-west filament.  On the other hand, as further discussed in Sect.~\ref{sec:MACS1447}, the X-ray cavities are co-spatial with a jetted radio outflow detected with the JVLA, which likely inflates the ICM and moderates the cooling and condensation of gas along the north-south direction. Indeed, the regions to the north and south of the BCG lack molecular gas emission, which instead preferentially extend along the east-west direction. Noteworthy, similar X-ray cavities have been previously observed in nearby clusters 
\citep{McNamara2001,Blanton2009,Walker2014,Su2017,HitomiColl2018}. {These results support} the scenario that radio-mode AGN feedback regulates gas cooling and shape the molecular gas morphology for MACS~1447, {similarly to RX~1532 (see Sect.~\ref{sec:RX1532_discussion}).}

{Further inspection of moment-0 and maxima maps reported in Fig.~\ref{fig:NOEMAmaps} reveals} that the molecular gas distribution and kinematics for MACS~1447 is complex and asymmetric. The molecular gas reservoir is globally extended along the east-west direction, with three distinct CO(3$\rightarrow$2) peaks associated with the E, 0, and W components. Interestingly, the component E has its peak close to the BCG center coordinates, and does not cover the full extent of the BCG stellar emission towards the east, while instead the molecular gas components 0 and W cover the full extent of the filamentary structure, which extends towards south-west (see Fig.~\ref{fig:allBCGs_optical_CO}). Similarly, the velocity (moment-1) map is complex, and shows clear discontinuities of the order of $\sim\pm100$~km~s$^{-1}$ along the east-west direction, particularly between the components E and 0 as well as between the components 0 and W. Furthermore, both E and 0 components, which have the highest CO(3$\rightarrow$2) peak values, also show the highest velocity dispersion of $\sim200$~km~s$^{-1}$, that is, higher than $\sim100$~km~s$^{-1}$  of the molecular gas filament. The components 0 and W are closer to the BCG core than the filament (component E). For the former the gravitational potential is thus deeper, and hence the velocity dispersion is higher. 

Altogether, the asymmetric elongated morphology and velocity discontinuities of the molecular gas reservoir of MACS~1447 support a scenario where the molecular gas is associated with the BCG core and an associated tail of ram pressure stripped gas. For a comparison, similar extended gas morphology has been previously found in some spectacular cases of ram pressure stripping in clusters such as ESO~137-001 in the nearby Norma cluster \citep{Jachym2014,Jachym2019} and the southern companion of the SpARCS~1049+56 BCG at $z = 1.709$ \citep{Castignani2020b}. {However, for SpARCS~1049+56, an alternative run-away cooling scenario has been also proposed \citep{Hlavacek-Larrondo2020,Barfety2022}.}

{Concerning MACS~1447,} the motion of the BCG is likely occurring mostly along the plane of the sky towards east. Line of sight velocities are moderate; that is, they have absolute values of $\lesssim200$~km~s$^{-1}$. Ram pressure stripping is expected to be effective in processing the gas in cluster core galaxies, as the net pressure is proportional to both ICM density and the square of the galaxy velocity relative to the ICM. Cool-core (low-entropy) cluster cores have high ICM densities, with number densities up to $\sim1$~electrons~cm$^{-3}$, and high galaxy velocity dispersions, up to $\sim1000$~km~s$^{-1}$ and beyond \citep[e.g.][for some reviews]{Dekel_Ostriker1999,Boehringer_Werner2009}.

The BCG motion in the dense ICM of the cluster core has likely given origin to the long $\sim25$~kpc tail of molecular gas (i.e., components 0 and W), as well as to the compression of gas at the shock front, possibly close to or coincident with the CO(3$\rightarrow$2) emission peak of the component E. When infalling velocities are high, ram pressure tails can extend for tens of kiloparsecs outward from the galaxy \citep{Sheen2017,Russell2017,Jachym2019}.

{Interestingly, the MACS~1447 cluster is part of the ongoing Kaleidoscope Survey with MUSE of over 180 clusters \citep{Jauzac2019,Lagattuta2023,Patel2024}
 and was observed in March 2019 (ID:~0102.A-0718; PI:~A.~Edge). The optical line emission is strongly asymmetric and filamentary (Halson, in prep). By unveiling the morphology, content, and kinematics of the ionized gas in and around the BCG, these MUSE observations can also potentially confirm the presence of a shock front at the location of the component E.}

{Similarly to ram pressure stripping, the observed phenomenology could be a result of sloshing motions in the ICM \citep{Vantyghem2019}, which often occur in disturbed clusters such as mergers. Interestingly, \citet{PrasowEmond2020} reported some pieces of evidence for the presence of sloshing motions in the cluster: strong metallicity gradients, the plume-like structure in the X-rays, the optical filaments, and the strong ellipticity of the cluster tend to favor a past merger event, and thus sloshing motions. It is difficult, however, to discriminate between the two physical mechanisms, as both ram pressure stripping and ICM sloshing imply a relative motion of the BCG with respect to the ICM and ultimately result in similar environmental processing in terms of the observed extended BCG morphology.} 
{In summary, to explain the overall observed phenomenology of MACS~1447, we propose two possible scenarios:  ram pressure stripping or sloshing motions, or a combination of the two.}

\subsection{CHIPS~1911: Tidal tails}

The color-composite image of CHIPS~1911 in Fig.~\ref{fig:allBCGs_optical_CO}, taken from \citep{Somboonpanyakul2021a}, displays an elongated stellar morphology associated with the BCG complex, characterized by an optically bright host galaxy and two tails, extended towards the north and the south; the latter brighter than the former. The two tails appear to have blue colors, and may thus be associated with ionized gas. To test this hypothesis,  similarly to the other two BCGs considered in this work, we retrieved publicly available {\it HST } images from the HLA archive. Filamentary structures elongated along the north-south direction are visible in the F550M ACS and F110W WFC-IR images, independently, {which are used in the color-composite image in Fig.~\ref{fig:allBCGs_optical_CO}, and} are centered around the redshifted [O~II] and H$\alpha$ lines, respectively.

Furthermore, as illustrated by the CO(2$\rightarrow$1) contours that are overlaid in Fig.~\ref{fig:allBCGs_optical_CO} (solid green lines),  the bulk of molecular gas reservoir (component 0, see Fig.~\ref{fig:NOEMAmaps}) is co-spatial with the stellar emission of the BCG. We also clearly detect an extended molecular gas emission towards the north (component N), co-spatial with the stellar stream extending to the north of the BCG.  The observed molecular gas emission of component N could be associated with star formation within it, possibly associated with a tidal tail. The presence of two (quasi-)symmetric tails in the {\it HST} images further support the scenario of a tidal interactions. 

{It is worth mentioning that a comparison of the spatial extent and distribution of the cold molecular gas reservoir in CHIPS~1911 with respect to that of the other two BCGs of this work deserves caution.  We stress that we observed CO(2$\rightarrow$1) in CHIPS~1911, with a resolution of $\simeq2.5^{\prime\prime}$, which corresponds to $\simeq15$~kpc at the BCG redshift (see Sect.~\ref{sec:NOEMAobservations}). Higher frequency NOEMA CO(3$\rightarrow$2) observations for RX~1532 and MACS~1447 have instead a better resolution of $\simeq(0.7-0.8)^{\prime\prime}$, equals to $\simeq4$~kpc at the BCG redshifts.}

{Nonetheless,} the presence of strong velocity gradients along the north-south direction (see Fig.~\ref{fig:NOEMAmaps}) further strengthens the presence of tidal interactions {associated with CHIPS~1911} \citep[e.g.,][]{Weilbacher2002,Bournaud2004}. Similarly, the BCG is a starburst galaxy \citep{Somboonpanyakul2021a}, with the highest SFR among the three BCGs in our sample, exceeding 100~$M_\odot$/yr, while enhancements of star formation and starbursts are commonly associated with galaxy interactions and mergers \citep[e.g.,][]{Sanders_Mirabel1996,Ellison2013,Renaud2022}.
The presence of two symmetric extended tails favors the scenario of tidal interactions against ram pressure stripping or sloshing motions. 

Finally, it is intriguing that the molecular gas component N is co-spatial not only with the northern stellar stream seen by  {\it HST}, but also with a secondary low-entropy peak in X-rays. We refer to Fig.~\ref{fig:allBCGs_optical_CO}, right panel, where white contours denote the X-ray emission from {\it Chandra} \citep{Somboonpanyakul2021a}. Therefore, there exists the alternative possibility that the observed CO(2$\rightarrow$1) emission is originated by run-away cooling of the ICM. This interpretation is in agreement with that argued by \citet{Somboonpanyakul2021a}, who proposed that the young stars are forming directly from the cooling ICM.
However, unambiguous observations of runaway cooling in clusters are debated in the literature \citep{Su2023,Castignani2020b,Barfety2022}.

\section{Conclusions}
\label{sec:conclusions} 

We have presented new observations of the molecular gas reservoirs associated with three intermediate redshift ($z\simeq0.4$) BCGs from the NOEMA interferometer. We selected three of the most star-forming BCGs (${\rm SFR}\simeq100~M_\odot$/yr) {at intermediate redshifts} residing at the center of confirmed cool-core clusters. Namely, {the selected BCGs} are RX~1532 ($z=0.361$), MACS~1447 ($z=0.376$), and CHIPS~1911 ($z=0.485$). {These sources are thus} in a phase of rapid mass assembly, which makes them ideal targets for high-resolution observations of their molecular gas feeding star formation in order to unveil the physical mechanisms responsible in shaping the mass assembly and evolution of some of the most massive galaxies in the distant Universe (i.e., $M_\star\simeq(1-6)\times10^{11}~M_\odot$).

All three BCGs exhibit point-like millimeter continuum emission with a globally steep radio spectrum  (spectral slope $\alpha = 0.6$), which we interpret as being optically thin synchrotron emission from the AGN. The three BCGs have low-frequency radio luminosities typical of FR~I radio galaxies in the range $L_{\rm 1.4~GHz}\simeq(0.9-1.9)\times10^{25}$~W~Hz$^{-1}$, which is similar to famous local BCGs such as M87 or Perseus~A.

We performed an analysis of NOEMA intensity and velocity maps in CO(2$\rightarrow$1) for CHIPS~1911 and in CO(3$\rightarrow$2) for RX~1532 and MACS~1447. We then combined them with archival deep optical/infrared images from the {\it Hubble Space Telescope} that trace stellar and ionized gas emission.
We unambiguously detected in situ star formation, filaments of accreting cold gas likely regulated by AGN feedback, a disturbed morphology associated with tidal tails of molecular gas, and gas compression and tails originating from ram pressure stripping or {sloshing motions.}

{Altogether, the three BCGs are unambiguously associated with large ($M_{H_2}\simeq10^{11}~M_\odot$) and extended (over tens of kiloparsecs) molecular gas reservoirs with complex morphologies and kinematics. The condensation of gas onto the BCGs, modulated by AGN feedback and associated cycles, has ultimately resulted in the observed molecular gas reservoirs. However, they exhibit significant differences in terms of distribution and kinematics. Namely, we found flows of cooling gas (RX~1532), ram pressure stripping or sloshing motions (MACS~1447), and galactic tides (CHIPS~1911).}

{Although the sample size is limited and thus prevents us from drawing firm conclusions, the observed variety of environment-driven mechanisms responsible for the processing of cold and ionized gas is intriguing and deserves further investigation with larger samples of intermediate-$z$ BCGs. It may also be a selection bias,} as we selected high star forming systems, thus privileging BCGs caught in a peculiar starbursting phase of their evolution.  Altogether, our study provides new insights into the physical mechanisms responsible for the mass assembly of {BCGs at the center of intermediate-$z$ clusters.}

\begin{acknowledgements}
{We thank the anonymous referee for helpful comments which contributed to improve the paper.}
 This work is based on observations with the IRAM~30m telescope, project number 274-10 (PI: A.~Edge), and with the IRAM NOEMA Interferometer, carried out under projects S21BX (PI:~Combes), W21CL (PI:~Salomé) and S23BY (PI:~Castignani). IRAM is supported by INSU/CNRS (France), MPG (Germany) and IGN (Spain). GC  acknowledges the support from the Next Generation EU funds within the National Recovery and Resilience Plan (PNRR), Mission 4 - Education and Research, Component 2 - From Research to Business (M4C2), Investment Line 3.1 - Strengthening and creation of Research Infrastructures, Project IR0000012 – “CTA+ - Cherenkov Telescope Array Plus”. {AE acknowledges support from STFC grant ST/T000244/1.}  This work is partly based on tools and data products produced by GAZPAR operated by CeSAM-LAM and IAP. GC thanks Denis Burgarella and Olivier Ilbert for useful discussion about GAZPAR data products. 
\end{acknowledgements}
\bibliographystyle{aa}
\bibliography{paper}

\newpage

\label{lastpage}
\onecolumn

\begin{appendix}

\section{NOEMA spectra of the BCG sub-regions}\label{sec:NOEMAspectra_subregions}
In this section, we provide CO spectra for the three BCGs corresponding to the sub-regions defined in Fig.~\ref{fig:NOEMAmaps}.

\begin{figure*}[h!]
\centering
\includegraphics[width=0.8\textwidth]{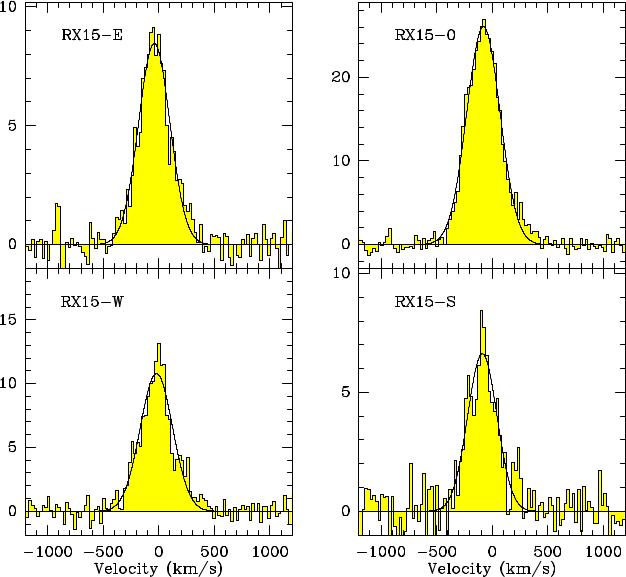}
\caption{CO(3$\rightarrow$2) spectra for RX~1532 sub-regions E, 0, W, and S defined as in Figure~\ref{fig:NOEMAmaps}. The solid lines show the Gaussian best ﬁts. In each spectrum, the ﬂux (y-axis,) in units of mJy, is plotted against the relative velocity with respect to the BCG redshift, as in Table~\ref{tab:BCG_properties}.}
\label{fig:RX1532_COspectra_regions}
\end{figure*}

\begin{figure*}[h!]
\centering
\includegraphics[width=0.8\textwidth]{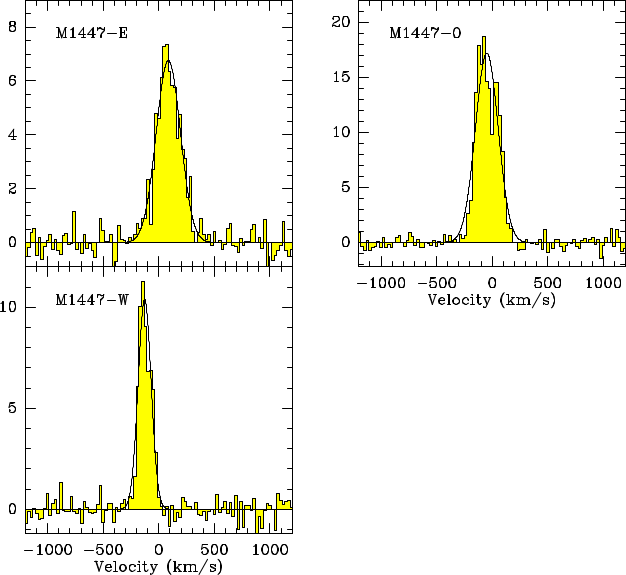}
\caption{CO(3$\rightarrow$2) spectra for MACS~1447 sub-regions E, 0, and W defined as in Figure~\ref{fig:NOEMAmaps}. The solid lines show the Gaussian best ﬁts. In each spectrum, the ﬂux (y-axis,) in units of mJy, is plotted against the relative velocity with respect to the BCG redshift, as in Table~\ref{tab:BCG_properties}.}
\label{fig:MACS1447_COspectra_regions}
\end{figure*}

\begin{figure*}[h!]
\centering
\includegraphics[width=0.8\textwidth]{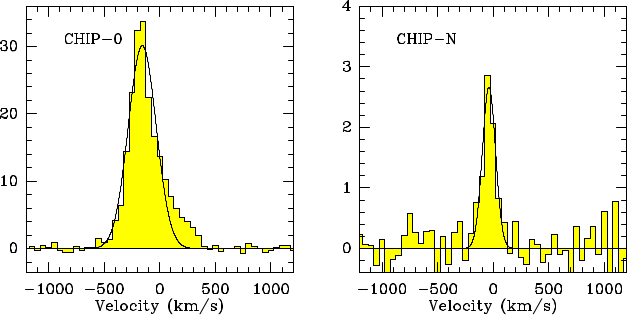}
\caption{CO(2$\rightarrow$1) spectra for CHIPS~1911 sub-regions 0 and N defined as in Figure~\ref{fig:NOEMAmaps}. The solid lines show the Gaussian best ﬁts. In each spectrum, the ﬂux (y-axis,) in units of mJy, is plotted against the relative velocity with respect to the BCG redshift, as in Table~\ref{tab:BCG_properties}.}
\label{fig:CHIPS1911_COspectra_regions}
\end{figure*}

\end{appendix}

\end{document}